\documentclass[a4paper,10pt]{article}
\usepackage[margin=1.95cm]{geometry}
\usepackage[T1,T2A]{fontenc}
\usepackage[utf8]{inputenc}
\usepackage{authblk}
\usepackage{lipsum}
\usepackage{amssymb,amsthm}
\usepackage{amsmath}
\usepackage{amscd}
\usepackage{mathrsfs}
\usepackage{accents}
\usepackage{listings}
\usepackage{graphicx}
\usepackage[all]{xy}
\usepackage{enumitem}
\usepackage{changepage}
\usepackage{subcaption}
\usepackage[linecolor=green,backgroundcolor=green,bordercolor=green]{todonotes}
\usepackage{graphicx}
\usepackage{float} 
\usepackage{algorithm2e}
\usepackage{algpseudocode}
\usepackage{chemist}
\usepackage[version=3]{mhchem}
\usepackage{diagbox}
\usepackage{framed} 
\usepackage{lscape}
\usepackage{array, boldline, makecell, booktabs}
\usepackage{mathtools}
\usepackage{multicol}
\usepackage{color}
\usepackage{hyperref,tikz}
\usepackage{pdfpages}

\usepackage{comment}

\definecolor{lime}{HTML}{A6CE39}
\DeclareRobustCommand{\orcidicon}{
	\begin{tikzpicture}
	\draw[lime, fill=lime] (0,0) 
	circle [radius=0.16] 
	node[white] {{\fontfamily{qag}\selectfont \tiny ID}};
	\draw[white, fill=white] (-0.0625,0.095) 
	circle [radius=0.007];
	\end{tikzpicture}
	\hspace{-2mm}
}

\foreach \x in {A, ..., Z}{\expandafter\xdef\csname orcid\x\endcsname{\noexpand\href{https://orcid.org/\csname orcidauthor\x\endcsname}
			{\noexpand\orcidicon}}
}

\makeatletter

\newtheorem*{rep@theorem}{\rep@title}
\newcommand{\newreptheorem}[2]{%
\newenvironment{rep#1}[1]{%
 \def\rep@title{#2 \ref{##1}}%
 \begin{rep@theorem}}%
 {\end{rep@theorem}}}
\makeatother

\newtheorem{theorem}{Theorem}
\numberwithin{theorem}{section}

\newtheorem{lemma}[theorem]{Lemma}
\newtheorem{corollary}[theorem]{Corollary}
\newtheorem{definition}[theorem]{Definition}

\newtheorem{remark}[theorem]{Remark}
\newtheorem{example}[theorem]{Example}

\newreptheorem{theorem}{Theorem}
\newreptheorem{lemma}{Lemma}
\newreptheorem{corollary}{Corollary}

\newcommand{\RR}{\mathbb{R}}
\newcommand{\QQ}{\mathbb{Q}}

\newcommand{\CC}{\mathbb{C}}
\newcommand{\ZZ}{\mathbb{Z}}
\newcommand{\isEquivTo}[1]{\underset{#1}{\sim}}

\begin{document}

\title{Algebraic study of  receptor-ligand systems:
a dose-response analysis}

\author[$1$]{L\'ea Sta\orcidA{}}
\author[$2,3$]{Michael Adamer\orcidB{}}
\author[$1,4, *$]{Carmen Molina-Par\'is\orcidC{}}

\affil[$1$]{ \small School of Mathematics, University of Leeds,
Leeds LS2 9JT, UK}

\affil[$2$]{ \small Department of Biosystems Science and Engineering,
ETH Z{\"u}rich, Mattenstrasse 26, 4058 Basel,
Switzerland}

\affil[$3$]{ \small Swiss Institute of Bioinformatics,
Quartier Sorge, Batiment Amphipole,
1015 Lausanne,
Switzerland
}

\affil[$4$]{ \small Theoretical Biology and Biophysics,
Theoretical Division,
Los Alamos National Laboratory,
Los Alamos 87545, NM, USA}
\affil[$*$]{ \small Corresponding author: molina-paris@lanl.gov}

\date{23 de Junio del 2022}

\providecommand{\keywords}[1]
{
  \small	
  \textbf{\textit{Keywords --}} #1
}


\maketitle

\begin{abstract}

The  study of a receptor-ligand system 
generally relies on the analysis of its dose-response (or concentration-effect) curve, which 
quantifies the relation between 
 ligand concentration and the biological effect  (or cellular response) induced when binding
 its specific cell surface receptor.
 Mathematical models of receptor-ligand systems have been developed
to compute a dose-response curve under the assumption that
 the biological effect  is proportional to the number of ligand-bound
 receptors. 
 Given a dose-response curve, two quantities (or metrics) have been defined
 to characterise the properties of the ligand-receptor system under consideration:
 amplitude and 
  potency (or half-maximal effective concentration, and denoted by EC$_{50}$).
  Both the amplitude and the EC$_{50}$  are 
    key quantities commonly used in pharmaco-dynamic modelling,
    yet a comprehensive mathematical investigation of the behaviour of these
    two metrics is still outstanding; 
    for a large  (and important) family
    of receptors, called cytokine receptors,
 we still do not know how amplitude and EC$_{50}$ 
    depend on receptor copy numbers.
 Here we make use of algebraic approaches (Gr\"obner basis) to study these
metrics
 for a large class of receptor-ligand models,
with a focus on cytokine receptors.
 In particular, we introduce a method,
making use of two motivating examples based on the 
interleukin-7 (IL-7) receptor, to compute analytic expressions for the amplitude and the EC$_{50}$. 
We  then extend the method 
to a wider class of 
receptor-ligand systems, sequential receptor-ligand
systems with extrinsic kinase, and  provide some 
examples.
The algebraic methods developed in this paper not only 
reduce computational costs and numerical errors, 
but allow us to explicitly 
identify key molecular parameter and rates which determine
the behaviour of the
dose-response curve. Thus, the proposed methods 
provide a novel and useful 
approach to perform
 model validation, assay design and parameter exploration of receptor-ligand
 systems.

\end{abstract}

\keywords{dose-response, cytokine, receptor,
Gr\"obner basis, amplitude, half-maximal effective concentration, steady state}

\section{Introduction}
\label{sec:introduction}

The human body consists of more than $3\times 10^{13}$ cells~\cite{bianconi2013estimation}, each of them receiving,
at any given time,
hundreds of signals 
from 
extra-cellular  molecules
when these bind their specific membrane receptors. These signals are integrated, translated
and read by a small number of intra-cellular molecules to generate appropriate cellular responses~\cite{janes2013models}.
Surface receptors specifically bind to extra-cellular molecules
called ligands.
The binding of a ligand to its receptor induces an intra-cellular cascade
of signalling events which regulate a cell’s fate,
such as migration, proliferation, death, or differentiation~\cite{maxwell2008receptor,uings2000cell}.
Receptor-ligand interactions are essential
in cell-to-cell communication, as is the case for immune cell
populations~\cite{farhat2021modeling}, 
and thus, a large body of literature has been devoted to the experimental and theoretical study of  cell signalling dynamics~\cite{wiley2003computational,ring2012mechanistic,gonnord2018hierarchy,rochman2009new,leonard2019gammac,lauffenburger1996receptors,park2019il7,molina2013mathematical,feinerman2010single}. 
Exploiting the controlled environment of \textit{in vitro} experiments, most cell signalling studies focus on the estimation of the affinity constant 
for a given receptor-ligand system, and the quantification
of biochemical on and off rates for the binding and unbinding,
respectively,
of receptor and ligand molecules.  Recent single-cell studies
have shown that cells have heterogeneous expression levels of receptor copy numbers.
Not only does the copy number depend on the cell type,
but  receptor copy numbers 
vary strongly  between isogenic cells of one cell type~\cite{feinerman2010single,cotari2013cell,farhat2021modeling}.
Given the heterogeneity
of receptor copy numbers across  and within cell types, it is timely
to understand how  a cell's response
to a given ligand depends on the expression
levels of its receptor.
This quantification will be a first step to account 
for the variability of receptor expression levels when designing and studying receptor-ligand models (both from an experimental
and mathematical perspective)~\cite{farhat2021modeling,cotari2013cell,gonnord2018hierarchy,ring2012mechanistic}. 

The  study of a receptor-ligand system generally relies on the analysis of its dose-response (or concentration-effect) curve, which 
describes the relation between 
 ligand concentration and the biological effect  (or cellular response) it generates when binding
 its specific receptor~\cite{maxwell2008receptor,lambert2004drugs,lauffenburger1996receptors}. 
 Mathematical models of receptor-ligand systems have been developed
to compute a dose-response curve, under the assumption that
 a biological effect is proportional to the number of ligand-bound
 receptors~\cite{molina2013mathematical,cotari2013cell,farhat2021modeling,wiley2003computational}. 
 Given a dose-response curve, two quantities (or metrics) have been defined
 to characterise the properties of the ligand-receptor system under consideration.
 These metrics are:
  the amplitude, which is defined as the difference between the maximal
  and minimal response, and the half-maximal effective concentration (or EC$_{50}$), which is the concentration of ligand required to induce an effect corresponding to 50\% of the amplitude~\cite{maxwell2008receptor,lambert2004drugs,lauffenburger1996receptors}.   
  The amplitude is a measure of the efficacy of the ligand,
  and the EC$_{50}$, a measure of the potency (or sensitivity) of the ligand (for a given receptor)~\cite{maxwell2008receptor,lambert2004drugs,dushek2011antigen}. 
  Both amplitude and EC$_{50}$  are 
    key quantities commonly used in pharmaco-dynamic modelling,
    yet a comprehensive mathematical investigation of the behaviour of these
    two metrics is still outstanding for most receptor-ligand systems.
    For instance, 
    for a large  (and important) family
    of receptors, called cytokine receptors~\cite{raeber2018role,villarino2015mechanisms,altan2019cytokine},
 we still do not know how amplitude and EC$_{50}$ 
    depend on receptor copy numbers (for a given concentration of ligand)~\cite{feinerman2010single,cotari2013cell}.
    
  In this paper 
 we bridge this gap by deriving closed-form expressions for a  class of cytokine-receptor models. We further highlight how tools from computational algebra can be used to facilitate the calculation of
 both the amplitude and the EC$_{50}$ for this family of models.
    
    Previous work has shown that the estimation of the amplitude and the EC$_{50}$ from experimental data is often possible, although strong inductive biases might be introduced~\cite{lambert2004drugs,maxwell2008receptor}. Usually one starts with a data set where the number (or concentration) of receptor-ligand signalling complexes formed (see Section~\ref{sec:amp-EC50}) is 
measured for different values of the  ligand concentration. Then, the estimation of the
amplitude and the EC$_{50}$
is turned into a regression problem by assuming a functional relationship in the data set and fitting a parametric curve. 
A simple first approach is to plot experimental values 
(corresponding to a measurable variable which quantifies cellular response)
as a function of ligand concentration.
 The amplitude and the EC$_{50}$ are then read directly from a curve formed by interpolation of the data points. Since the EC$_{50}$ is likely to fall between two data points, a geometrical method~\cite{alexander1999simple} can be used for an accurate determination. Nowadays many software packages can compute the amplitude and the EC$_{50}$  from the data set making use of statistical methods,
 which consist in finding the ``best-fit" equation to the dose-response curve. The most common shape of the
 dose-response curve is a sigmoid, and thus,  can be fitted with the famous Hill equation~\cite{gesztelyi2012hill,goutelle2008hill}.
However, other functions are also possible, such as a logistic equation~\cite{chen2013ec50,li2015comparison}, a log-logistic equation~\cite{jiang2014summarizing,suriyatem2017predictive}, or the Emax model~\cite{macdougall2006analysis,thomas2006hypothesis}. An asymmetrical sigmoid equation is sometimes needed for better precision~\cite{chen2013ec50,suriyatem2017predictive}. The amplitude and
the EC$_{50}$ are parameters of these equations and can thus,  be directly inferred from the fitting process. 
When a data set does not follow the strictly increasing pattern of these Hill-like functions, 
then more complex functions, such as bell-shaped curves~\cite{rovati1994lower}, or multi-phasic curves~\cite{di2015automated} can be used.
It is important to note that even though these empirical regression methods
allow one to quantify the two key receptor-ligand metrics, 
amplitude and EC$_{50}$,
they do not offer any 
mechanistic 
insights for the receptor-ligand system under consideration.
To this end, mathematical models can be used  to
describe the receptor-ligand system 
at a molecular
level; that is, mathematical models consider the biochemical reactions
which initiate a
cellular response~\cite{eftimie2016mathematical,wiley2003computational,lauffenburger1996receptors}. The challenge 
 in such  models is finding analytical, ideally closed-form, expressions for the amplitude and 
 the EC$_{50}$. 
 Due to the non-linear nature of the biochemical reactions involved, 
 this poses a significant and practical challenge.

Cytokine-receptor systems are of great relevance in immunology~\cite{raeber2018role,villarino2015mechanisms,altan2019cytokine},
and
 here
 we want to address this challenge in the context of this family of receptors~\cite{lin2018common,altan2019cytokine}.
 The advantages of having analytical (or closed-form) expressions 
 of the amplitude and the EC$_{50}$ for a large class
 of receptor-ligand systems are many: i) they allow
 to quantify their dependence on receptor copy numbers,
 ii) they facilitate mathematical model validation and parameter exploration,
 and iii) they reduce computational cost.
 To the best of our knowledge
 such expressions have been obtained in a few
 instances:
  closed or open bi-molecular receptor-ligand systems~\cite{gabrielsson2018lost}, monomeric receptors~\cite{mack2008exact}, or ternary complexes~\cite{douglass2013comprehensive}. More complicated receptor-ligand models have been studied with chemical reaction network theory (CRNT)~\cite{feinberg1987chemical,otero2017chemical,shiu2010algebraic,otero2017chemical}, 
  but  CNRT 
  has thus far, been focused on the analysis of the steady state of the system ({\em i.e.,} existence
  and number of steady states and their stability).
  Yet, we believe CRNT  is an essential and useful framework to start any  mathematical investigation of  the amplitude and  the EC$_{50}$.
  
Another aspect which can be  effectively addressed by mechanistic mathematical modelling is the effect of internal or external perturbations to the state of a cell. For example, in single-cell experiments or even repetitions of bulk experiments~\cite{cotari2013cell,feinerman2010single}, the experimental conditions can never be replicated exactly. This leads to noise not only in the measured quantities, but also in the reaction mechanisms themselves. 
This variation can be captured in mathematical models
which encode parameters such as affinity constants or total copy number of constituent molecular species. An analytical  study of the dependency of pharmacologically relevant quantities, such as amplitude and EC$_{50}$,
on the reaction parameters can facilitate \textit{in silico} drug design~\cite{moraga2017synthekines}.
While amplitude and EC$_{50}$ are widely employed to characterise biological phenomena, 
the manner in which they depend on the parameters of the receptor-ligand model  is not fully
understood. Thus, improved understanding of these relationships could 
provide novel biological and computational insights.

Motivated by the previous challenges and making use of 
methods from
CRNT and algebraic geometry, such as the Gr\"obner basis, in this paper we propose a new 
method to obtain analytic expressions of the amplitude and 
the EC$_{50}$ for a large class of receptor-ligand models,
with a focus on cytokine receptors.
The paper is organised as follows. In Section~\ref{sec:mathbackround} we introduce the mathematical background and essential notions of CRNT used in the following sections. With
IL-7
 cytokine receptor as a paradigm, in Section~\ref{sec:method} we propose a general method to calculate
the amplitude and the EC$_{50}$  of the dose-response curve for 
a class of receptor-ligand systems. 
In Section~\ref{sec:generalisation-SRLK}
we generalise the previous results 
to a wider class of 
receptor-ligand systems, sequential receptor-ligand
systems with extrinsic kinase, and  provide a few biological
examples of these systems.
Finally, we discuss and summarise our results in 
 Section~\ref{sec:conclusion}.
 We have included an appendix to provide additional details
 of our methods (perturbation theory) and our algebraic computations.


\section{Mathematical background} 
\label{sec:mathbackround}

In this section we briefly summarise the relevant notions of chemical reaction network theory and formally define amplitude, EC$_{50}$, and signalling function. A very short introduction to the use of Gr{\"o}bner bases is also given.


\subsection{A brief introduction to chemical reaction network theory}
\label{sec:CRNT}

In this paper  we view a chemical reaction network (CRN), $\mathcal{N}$, as a multi-set $\mathcal{N} = \{\mathcal{S},\mathcal{C},\mathcal{R}\}$, where $\mathcal{S}$ is the set of species, $\mathcal{C}$ the set of complexes, and $\mathcal{R}$ the set of reactions.
We note that in the context of CRN, a ``complex'' is a linear combination of species and need not be a ``biological functional unit'', which we refer to as a \emph{biological complex}. We denote, whenever useful, a biological complex formed by species $X$ and $Y$ as $X:Y$, where the colon denotes the physical bond between $X$ and $Y$. The order of species in the biological complex is irrelevant, {\em i.e.,} $X:Y$ = $Y:X$.

\begin{example}[Heterodimeric receptor tyrosine kinase]
  A simple heterodimeric receptor tyrosine kinase (RTK) model has a species set $\mathcal{S} = \{X_1,X_2,Y_1\}$, a complex set $\mathcal{C}=\{X_1+X_2,Y_1,Y_2\}$, and a reaction set $\mathcal{R}=\{X_1+X_2 \to Y_1, Y_1\to X_1+X_2, Y_1\to Y_2, Y_2\to Y_1\}$.
  Ligand binding induces dimerisation of these receptors resulting in auto-phosphorylation of their cytoplasmic domains (tyrosine autophosphorylation sites)~\cite{schlessinger2000cell}.
 $X_1$ and  $X_2$ are the two components of the heterodimeric RTK.
  The biological complexes  $Y_1 = X_1:X_2$
  and $Y_2=L:Y_1$ are the
  heterodimeric
  receptor with intrinsic kinase activity
  and the heterodimeric receptor bound
  to the ligand, respectively. In this paper the ligand concentration
 ($L$)
  is taken to be an input parameter and, hence, it does not feature as a separate chemical species in the species set $\mathcal{S}$.
\end{example}

We can associate a reaction graph to every CRN $\mathcal{N}$, by letting the vertex set be $\mathcal{C}$ and the (directed) edge set $\mathcal{R}$. There exists a class of important CRNs defined by their network reversibility.
\begin{definition}[Network reversibility]
Let $\mathcal{N}$ be a CRN with its associated reaction graph $\mathcal{G}(\mathcal{C},\mathcal{R})$.
An edge between $C_i$ and $C_j\in\mathcal{C}$ exists if $C_i\to C_j\in\mathcal{R}$. If for every edge $C_i\to C_j\in\mathcal{R}$, the edge $C_j\to C_i\in\mathcal{R}$ also exists, then the network is called \emph{reversible}. If for every edge, $C_i\to C_j\in\mathcal{R}$, a directed path exists going back from $C_j$ to $C_i$, then the network is called \emph{weakly reversible}. All reversible networks are also weakly reversible.
\end{definition}

A general reaction from complex $C_i$ to complex $C_j$ can be written as
\begin{equation}
  \sum_{k=1}^n \alpha_{ik} \; X_k\rightarrow \sum_{k=1}^n \alpha_{jk} \; X_k,
  \label{eq:GenericReac}
\end{equation}
where the sum is over the set of species $(X_1, X_2, \ldots, X_n)$,  and $\alpha_i=(\alpha_{i1},...,\alpha_{in})^T$ and $\alpha_j=(\alpha_{j1},...,\alpha_{jn})^T$ are non-negative integer vectors. The corresponding reaction vector is given by $r = \alpha_j - \alpha_i$.
For a CRN with $n$ species and $m$ reactions we can now define the $n\times m$ matrix of all reaction vectors,
$\Gamma$, such that $\Gamma = (r_1,\ldots , r_m)$. This matrix is called the {\em stoichiometric matrix}.


\begin{example}[Heterodimeric RTK continued]
The reaction graph of the heterodimeric RTK model is given by 
\begin{equation*}
X_1 + X_2 \rightleftharpoons Y_1 \rightleftharpoons Y_2.
\end{equation*}
The model is reversible with reaction vectors $r_1 = (-1,-1,1,0)^T$ and $r_2 = (1,1,-1,0)^T, r_3 = (0,0,-1,1)^T, r_4 = (0,0,1,-1)^T$.
\end{example}

To derive dynamical properties from the static description so far provided, we make use of 
the law of mass action kinetics~\cite{horn1972general}. First, we assign a  rate constant $k \in \RR_{>0}$ to 
each and every reaction in the network. Second, we denote the concentration of species $X_i$ by $x_i$.
With this notation, we then
associate a monomial to every complex $C_i = \sum_k \alpha_{ik} \; X_k$, as follows
\begin{equation}
  x^{\alpha_i} = x_1^{\alpha_{i1}}\cdots x_n^{\alpha_{in}},
  \label{eq:monomComp}
\end{equation}
where $n$ is the number of species in the network.
We define the reactant complex of a reaction as the complex on the left hand side of reaction~\eqref{eq:GenericReac}. The reaction rate of a reaction is the monomial of its reactant complex multiplied by the rate constant. The flux vector, $R(x)$, is the $m\times 1$ column vector of all reaction rates. The ordinary differential equations (ODEs) governing the dynamics of the reaction network are given by
\begin{equation}
  \frac{dx}{dt} = \Gamma \; R(x),
  \label{eq:DynSys}
\end{equation}
where $\Gamma$ is the stoichiometric matrix (defined above). 
We note that the reaction rate of the ${i}^\text{th}$ reactant complex is the $i^\text{th}$ row in $R(x)$, and
similarly, the stoichiometry of ${i}^\text{th}$ reaction is given by the $i^\text{th}$ column of $\Gamma$.

From \eqref{eq:DynSys} we can also deduce the conserved quantities of the reaction network. 
That is,  if a vector exists, $z \in \ZZ^n$, such that $d(z^Tx)/dt = z^T\Gamma R(x) = 0$, the quantity $z^Tx$ is conserved. Consequently, the left kernel of $\Gamma$ defines a basis for the space of conserved quantities. In
this way, conservations induce linear relations between the variables. Informally we say that a molecular species $X_i$ is conserved if its total number of molecules, $N_i$, is constant. $N_i$ is determined by the initial conditions.

\begin{example}[Heterodimeric RTK continued]
The dynamical system associated with the heterodimeric RTK model is given by
\begin{equation*}
    \frac{dx}{dt} = \frac{d}{dt}\begin{pmatrix}x_1\\ x_2\\ y_1\\ y_2\end{pmatrix} = \begin{pmatrix} -1 & 1 & 0 & 0\\ -1 & 1 &0 & 0\\ 1 & -1 & -1 & 1\\ 0 & 0 & 1 & -1\end{pmatrix}\begin{pmatrix} k_1x_1x_2\\ q_1y_1\\ k_2y_1 \\ q_2y_2\end{pmatrix} = \begin{pmatrix} -k_1x_1x_2 + q_1y_1 \\ -k_1x_1x_2 + q_1y_1\\ k_1x_1x_2 - (q_1+k_2-q_2)y_1\\ k_2y_1-q_2y_2\end{pmatrix},
\end{equation*}
with  $k_i$ as the reaction constants of the forward chemical reactions ($\rightharpoonup$) and $q_i$ the reaction constants of the backward reactions ($\leftharpoondown$). A basis for the conservation equations is given by 
the linear relations $X_1 + Y_1 + Y_2 = N_1$ and $X_2 + Y_1 + Y_2 = N_2$.
These imply that the total amount of the species $X_1$ ($X_2$) is conserved by adding the amounts of the bound states of the molecule ($Y_1$ and $Y_2$) to the amount of free molecule $X_1$ ($X_2$).
\end{example}

We can now define the biologically relevant  steady states of a CRN.

\begin{definition}
    A vector $x^*$ is a biologically relevant steady state if $\Gamma R(x^*) = 0$ and $x^*_i > 0\; \forall i \in \{1,\ldots,n\}$.
\end{definition}
A useful connection between the static network structure (defined earlier) and the existence (and stability) of unique biologically relevant
steady states can be made via deficiency theory~\cite{feinberg1987chemical}.

\begin{definition}[Deficiency]
Let $\mathcal{N}$ be a CRN with $\ell$ connected components in the reaction graph and $\eta = \dim{\text{span}(r_1,\ldots,r_m)}$ be the dimension of the span of the reaction vectors. The deficiency of $\mathcal{N}$ is  then given by $$\delta = |\mathcal{C}|-\ell - \eta.$$
\end{definition}
The notion of network deficiency leads to one of the fundamental theorems of CRNT, the \emph{Deficiency Zero Theorem}~\cite{feinberg1987chemical}, which connects the network structure to the dynamics of a CRN.

\begin{theorem}[Deficiency zero theorem]
  Let $\mathcal{N}$ be a weakly reversible CRN with $\delta = 0$. Then the network has a unique biologically relevant
  steady state for every set of initial conditions, and this steady state is asymptotically stable.
  \label{thm:DefZero}
\end{theorem}

With certain additional conditions on the reaction rates (see Refs.~\cite{feinberg1989necessary,dickenstein2011far}),
biologically  relevant steady states are detailed balanced. This means that for every reaction of the form \eqref{eq:GenericReac}, the steady states satisfy
\begin{align*}
    K x^{\alpha_i} = x^{\alpha_j},
\end{align*}
where $K = k/q$, the ratio of the rate constants of the forward and backward reactions, is called the \emph{affinity constant}
of the reaction. 

\begin{example}[Heterodimeric RTK continued]
The heterodimeric RTK model has $3$ complexes, $1$ connected component and the dimension of the span of the reaction vectors is $2$;
hence, $\delta = 3 - 1 - 2 = 0$. Since the network is reversible, we know from Theorem~\ref{thm:DefZero} that there exists exactly one stable positive steady state for each set of initial conditions. One can show that in fact $y_1 = (k_1/q_1)x_1x_2$ and $y_2 = (k_2/q_2) y_1$.
\end{example}


\subsection{Signalling function: amplitude and half-maximal effective concentration}
\label{sec:amp-EC50}

In this paper we want to closely investigate  pharmacological properties of  receptor-ligand systems, rather than the steady state structure of the models. In particular,  we want to study 
the  dose-response (or concentration-effect) curve of the system, which 
describes the relation between 
 ligand concentration and the biological effect  (or cellular response) it generates when binding
 its specific cell surface receptor. 
 As mentioned in Section~\ref{sec:introduction}
a lot of effort has been devoted to explore the steady state structure 
of chemical reaction networks. In this paper
we make use of algebraic methods to explore the dose-response
of receptor-ligand systems. To do so we 
start with the definition of  signalling complex.
We note that in most biological
instances the 
signalling complex
is formed by all the 
sub-unit chains that make
up the full receptor,
 intra-cellular kinases and the specific ligand~\cite{janes2013models,uings2000cell,gonnord2018hierarchy,leonard2019gammac,feinerman2010single,cotari2013cell,dushek2011antigen}. 

\begin{definition}
The signalling complex of a receptor-ligand system is the biological complex which induces a biological response.
\end{definition}

This leads to the following definitions of the signalling function and dose-response curve.

\begin{figure}[htp!]
    \centering
        \includegraphics[width=\textwidth]{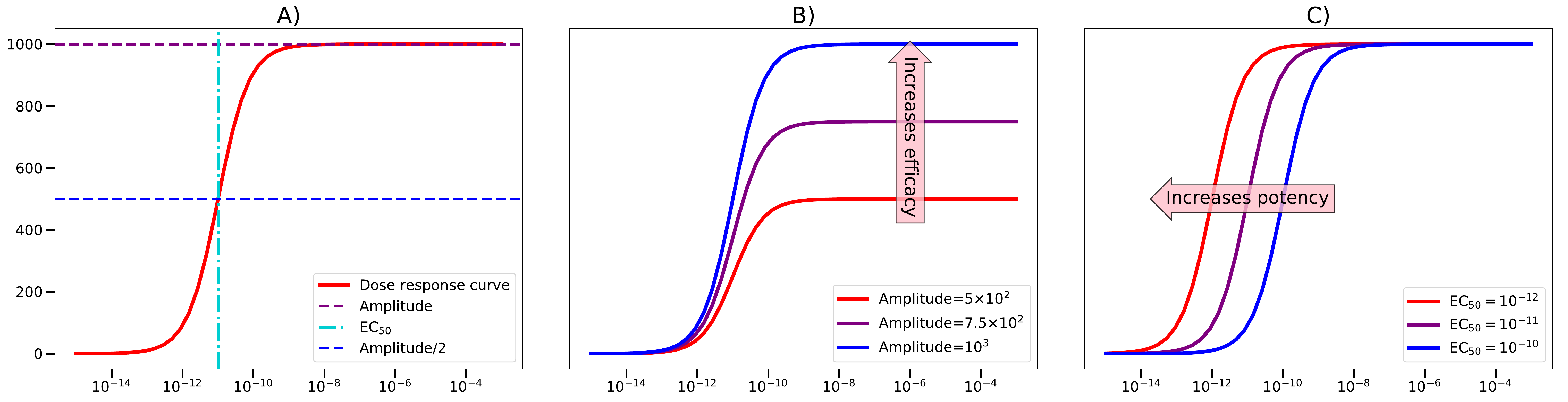}
    \caption{Sigmoid dose-response curve: number of signalling complexes formed, $\sigma(L)$, as a function of the concentration of ligand $L$ (arbitrary units). A) The maximum value defines the amplitude. The EC$_{50}$ is the concentration of ligand which corresponds to half the amplitude. B) Three dose-response curves with the same EC$_{50}$ value and different amplitudes. Increasing the amplitude shifts up the maximum of the curve and increases the efficacy. C) Three dose-response curves with the same amplitude and different EC$_{50}$ values. Decreasing the EC$_{50}$ shifts the  dose-response curve to the left and increases the potency of the ligand.}
    \label{fig:doseresponsecurve}
\end{figure}

\begin{definition}
We define the \emph{signalling function}, $\sigma:\RR_+\to\RR_+,\, L\mapsto \sigma(L)$, as the univariate function 
which assigns to a given value of ligand concentration, $L$, the number  (or concentration) of signalling complexes formed at steady state. 
The dose-response curve is the corresponding plot of the signalling function. 
\label{def:signalling_function}
\end{definition}
We note that in what follows we will not distinguish between number (or concentration) of signalling
complexes since one can be obtained from the other if we know the volume of the system and Avogadro's number.

The specific choice of $\sigma$ will depend on the receptor-ligand system under consideration. In this paper we focus on the class of cytokine receptors and the signalling function will be defined in Section~\ref{sec:method}.
In our examples the signalling function will
be a product of the steady state values 
(numbers) of
sub-unit chains that make
up the full receptor,
 intra-cellular kinases,  affinity constants of the reactions involved, and ligand concentration. 
 This together with 
 equations \eqref{eq:monomComp} and \eqref{eq:DynSys}
 indicate
 that the signalling function will always be  algebraic. 
Next, we define a central object of study in this paper; namely, the amplitude of the signalling function, often referred as efficacy in the pharmacology literature~\cite{maxwell2008receptor}.

\begin{definition}
The \emph{amplitude} of the signalling function, $A$,  is the difference between the maximum and the minimum of 
$\sigma$; that is,
 $A\equiv max(\sigma)-min(\sigma)$. 
\end{definition}
We note that when $\min(\sigma)=0$, which is the case considered in this paper ($\min(\sigma)=\sigma(0)=0$), the amplitude is given by the maximum of the signalling function. If, in addition, the dose-response curve attains its maximum at large concentrations (for instance, when the dose-response curve is sigmoid), we have
\begin{equation}
    A=\lim_{L\to+\infty}\sigma(L).
    \label{eq:asymptotic_amplitude}
\end{equation}
The amplitude provides information about the magnitude of the intra-cellular response to the stimulus, $L$. The larger the amplitude is, the larger the response variability will be. The amplitude is always bounded by the number of molecules available. However, this bound is often
not tight~\cite{perez2015implicit}. To quantify the sensitivity of the model to the stimulus, {\em i.e.,} the potency of the ligand $L$, we introduce the \emph{half-maximal effective concentration}, EC$_{50}$.

\begin{definition}
The half-maximal effective concentration, or \emph{EC$_{50}$}, is the ligand concentration $L^*$ which satisfies $\sigma(L^*)=
min(\sigma) + \frac{
max(\sigma)-min(\sigma)}{2}=
min(\sigma) +
\frac{A}{2}$.
\label{def:ec50}
\end{definition}
We say that the EC$_{50}$ is inversely proportional to ligand potency;  namely, the lower the EC$_{50}$, the higher the potency of the ligand. Figure~\ref{fig:doseresponsecurve} illustrates the amplitude and the EC$_{50}$ of a sigmoid dose-response curve
(A) when its minimum is zero: increasing the amplitude shifts up the maximum of the curve and results in greater efficacy (B),
and decreasing the EC$_{50}$ shifts the dose-response curve to the left and increases the potency of the ligand (C).
We now review some algebraic and analytic tools which  will enable us to compute the EC$_{50}$ and
the amplitude. 


\subsection{Gr\"obner bases}
\label{sec:groebner}

Since we assume the law of mass action, the models studied in this paper are systems of polynomial equations, and thus, we can use the techniques developed in the field of computational algebra and algebraic geometry~\cite{cox1997ideals}. Such methods have also been successfully applied to many topics in chemical reaction network theory, see {\em e.g.,} Refs.~\cite{gross2016algebraic,sadeghimanesh2019grobner,dickenstein2019multistationarity}.
In particular,  we make use of Gr{\"o}bner bases. Informally speaking, a Gr{\"o}bner basis is a non-linear generalisation 
of the concept of a basis in linear algebra and, therefore when a Gr{\"o}bner basis for a polynomial system is calculated, many properties
of the system can be investigated, such as  the number of  solutions and  the dimensionality of the space of solutions. Strictly speaking, however, a Gr{\"o}bner basis is not a basis as it is not unique and it depends on the 
 lexicographical (lex) monomial ordering chosen.
 For more details we refer the reader to Ref.~\cite{cox1997ideals}.

A lex Gr{\"o}bner basis is a triangular polynomial system; that is, for a polynomial system (ideal) in $\QQ[x_1,\ldots ,x_n]$ we obtain a polynomial system of the form
\begin{equation}
    g_n(x_1,\ldots ,x_n) = g_{n-1}(x_1,\ldots ,x_{n-1}) = \dots = g_1(x_1) = 0.
\end{equation}
We note that when the solution space is positive dimensional, then $g_1,\ldots , g_n$ are identically zero.
For a given Gr{\"o}bner basis with zero-dimensional solution space we can now iteratively, and often numerically, solve the constituent polynomials to obtain the solutions (in $\CC^n$) for the polynomial system. We can also find all real and, further, positive solutions, if there are any~\cite{cox1997ideals}. 


\section{Methods: analytical study of receptor-ligand systems}
\label{sec:method}

In this section we first outline the computation of the analytic expressions of the steady state, amplitude and EC$_{50}$ 
for two IL-7 receptor  (IL-7R) models. These two examples then allow us to introduce a more general method to analytically compute the amplitude and the EC$_{50}$ of receptor-ligand systems under the following hypotheses:

\begin{enumerate}

    \item The system is in steady state.
    
    \item The ligand is in excess (we consider ligand concentration, $L$, as a parameter instead of a dynamic variable).
    
   \item A unique biologically relevant solution exists for any given set of rate constants and initial conditions.
   
\end{enumerate}
The IL-7R models  we have chosen are simple enough to illustrate our method, and thus, to derive analytic expressions for
the amplitude and the EC$_{50}$, yet complex enough to show its limitations. 

\subsection{Two motivating examples: IL-7 cytokine receptor as a paradigm}

\begin{figure}[htp!]
    \centering
          \begin{subfigure}{.3\textwidth}
    \includegraphics[width=\textwidth]{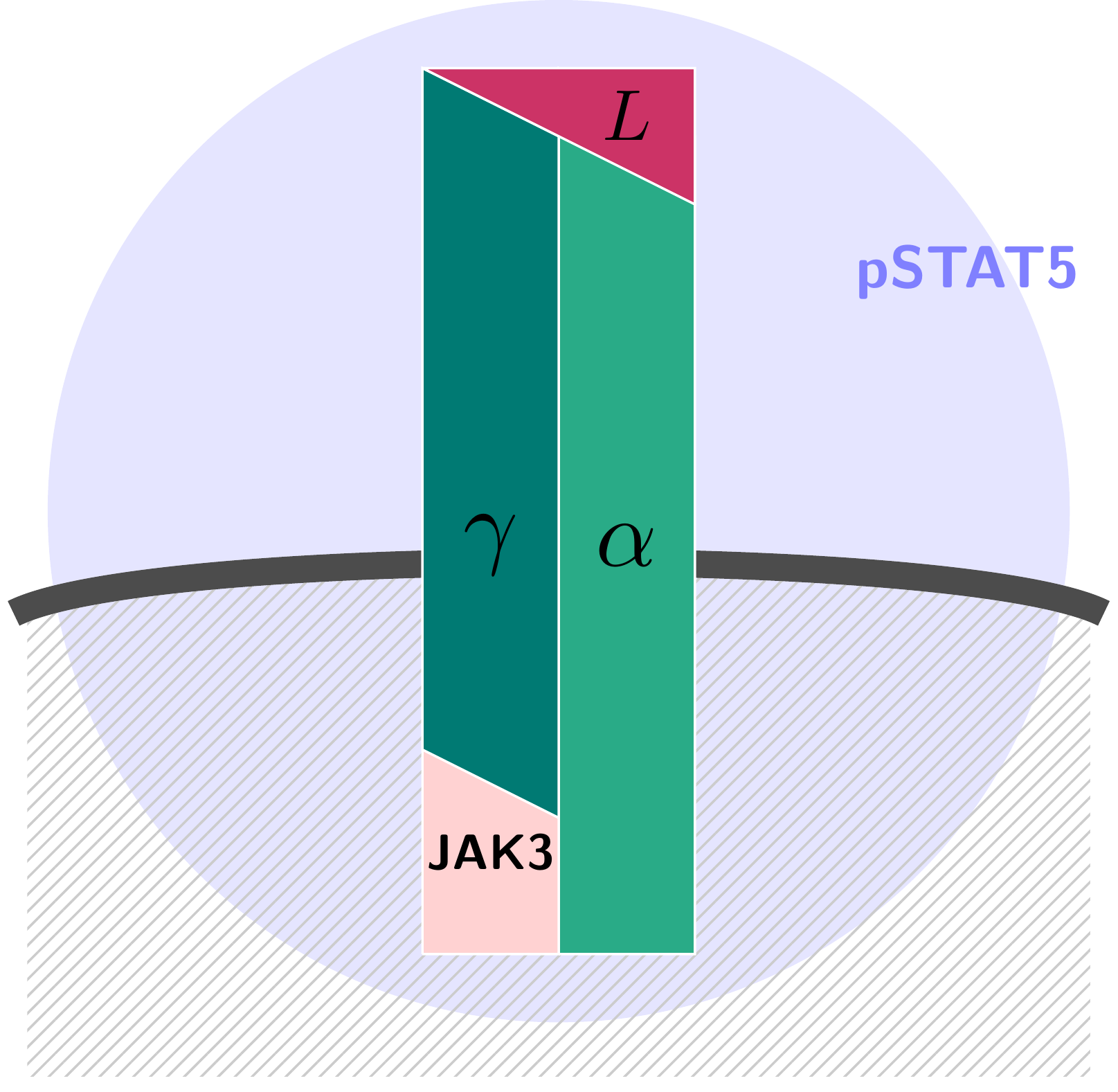}
    \caption{Signalling IL-7 complex.}
    \label{fig:IL7Rsig}
  \end{subfigure}
 \begin{subfigure}{.3\textwidth}
    \includegraphics[width=\textwidth]{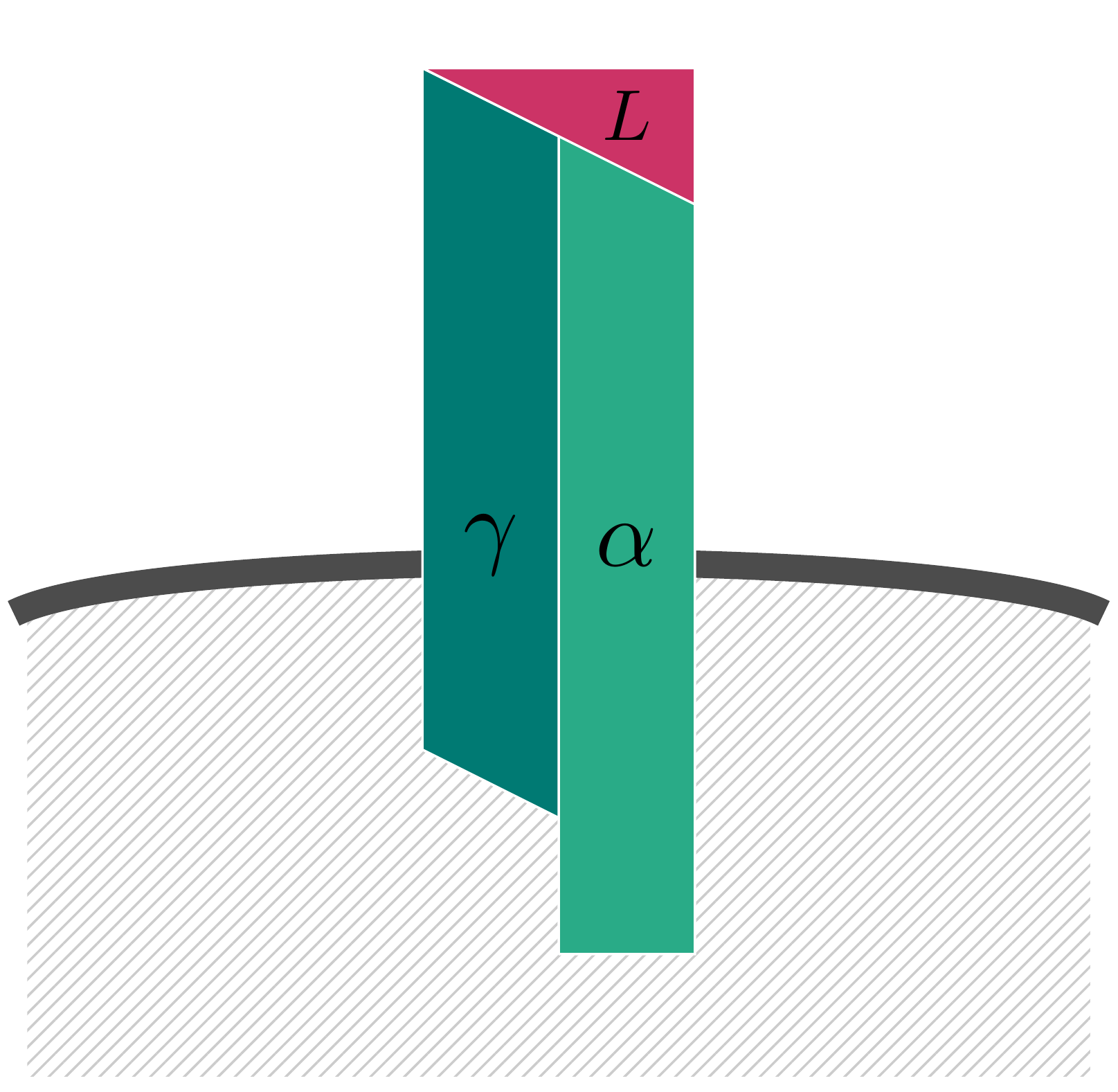}
    \caption{Dummy IL-7 complex.}
    \label{fig:IL7Rdum}
  \end{subfigure}
   \begin{subfigure}{\textwidth}
    \includegraphics[width=\textwidth]{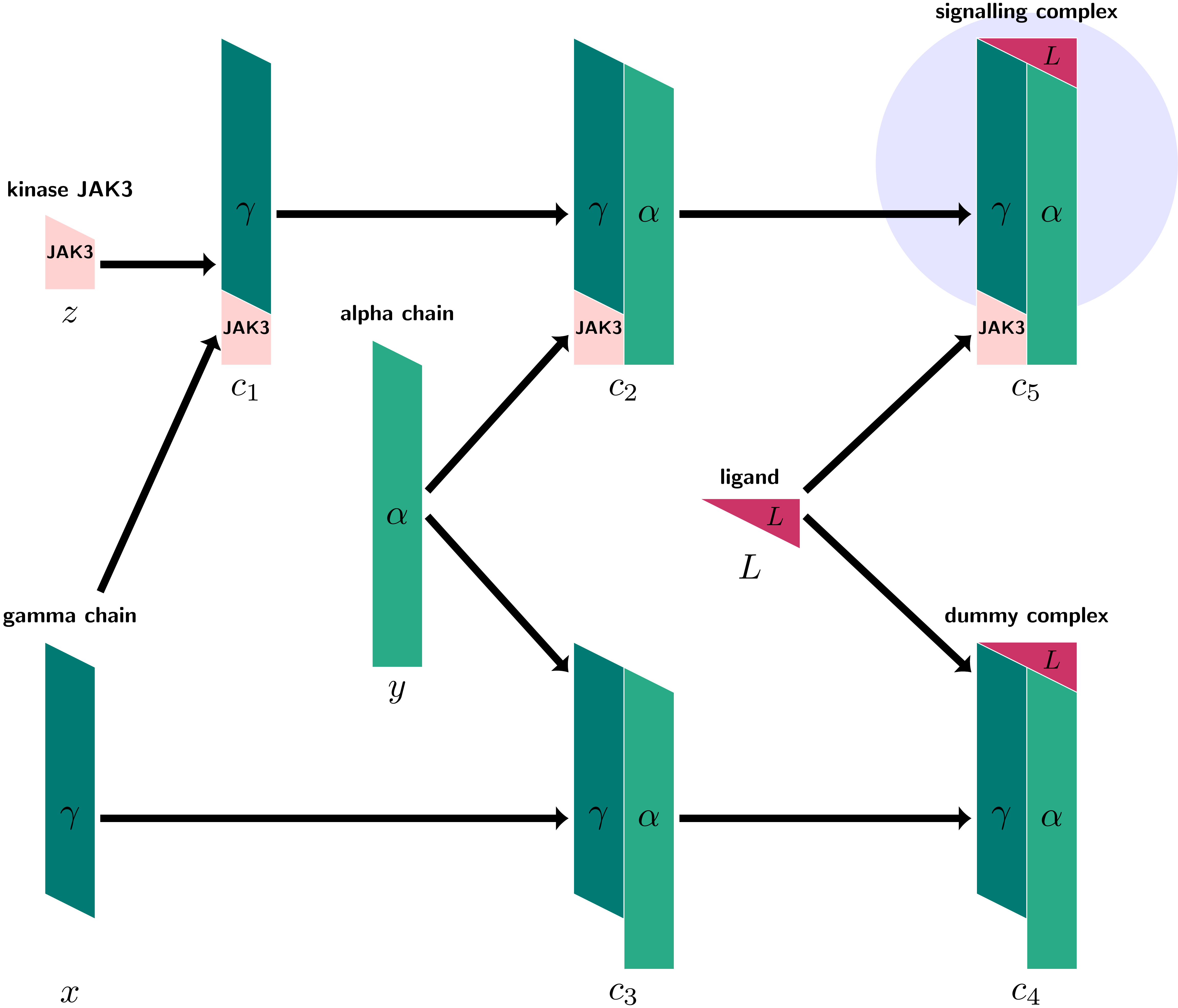}
    \caption{IL-7R model: sequential chemical reaction scheme.}
    \label{fig:IL7Rscheme}
  \end{subfigure}
    \caption{First IL-7R model: (a) The IL-7 receptor is composed of the trans-membrane $\gamma$ and $\alpha$ chains. The $\gamma$ chain can bind the intra-cellular downstream kinase JAK3. When the ligand, IL-7, binds the full receptor, it phosphorylates STAT5. (b) The IL-7R model allows the formation of ``dummy'' complexes: IL-7 bound IL-7R complexes, devoid of JAK3, which are unable to induce intra-cellular signalling. (c)  IL-7 bound IL-7R complexes with JAK3 are able 
    to induce intra-cellular signalling, and  thus, are called ``signalling'' complexes.
     IL-7R dummy and signalling complexes are formed sequentially. The mathematical notation used in this paper is shown below each molecule or complex.}
    \label{fig:IL7R}
\end{figure}

We now consider  the 
cytokine interleukin-7 (IL-7) and its receptor (IL-7R)~\cite{palmer2008interleukin, rochman2009new,gonnord2018hierarchy,leonard2019gammac,cotari2013cell,park2019il7} as a motivating
receptor-ligand system.
IL-7 is a cytokine involved in T~cell development, survival and homeostasis~\cite{ma2006diverse}. Its receptor, IL-7R, is displayed on the surface of T~cells and is composed of two trans-membrane chains: the common gamma chain (denoted by $\gamma$) and the specific high affinity chain IL-7R$\alpha$ (denoted by $\alpha$)~\cite{gonnord2018hierarchy,park2019il7,ma2006diverse,molina2013mathematical}.
Cytokine receptors do not contain 
  intrinsic kinase domains, and thus, make use of 
Janus family tyrosine kinases (JAKs) and signal in part by the activation of signal transducer
and activator of transcription (STAT) proteins~\cite{lin2019fine}.
In the case of the gamma chain, it binds to the intra-cellular extrinsic Janus kinase molecule, JAK3. Binding of IL-7 to the dimeric JAK3-bound IL-7 receptor, defined as $\alpha:\gamma:JAK3$, initiates a series of biochemical reactions from the membrane of the cell to its nucleus, which in turn lead to a  cellular response.
For the IL-7R system the STAT protein preferentially activated is STAT5~\cite{lin2019fine},
so that 
the amount of phosphorylated STAT5 can be used as 
the experimental measure of the intra-cellular response
generated by the IL-7 stimulus. The IL-7R receptor is illustrated in Figure~\ref{fig:IL7Rsig}, where the hatched area determines the intra-cellular environment. 

The first model  we  consider is shown in Figure~\ref{fig:IL7Rscheme}. As discussed in Ref.~\cite{rochman2009new}, the gamma chain is shared by other cytokine receptors. This model does not include the competition for the gamma chain between
different cytokine receptors, therefore later in this section we  introduce a second model to account for this
competition. 
In this section we will
provide an (algebraic) analytic treatment of both models.
We consider
the formation of ``dummy'' receptors, $\alpha:\gamma$, which are formed of the IL-7R devoid of JAK3 and, therefore, they cannot signal (see Figure~\ref{fig:IL7Rdum}). We further assume no allostery; that is, the affinity constants of the biochemical reactions involved in the formation of  the dummy complex, $L:\alpha:\gamma$, are the same as the affinity constants involved in the formation of  the signalling complex, $L:\alpha:\gamma:JAK3$.


\subsubsection{The IL-7 receptor-ligand system: two receptor chains and a kinase}
\label{sec:il7r}

We first consider a model in which the IL-7R is formed sequentially, 
one molecule at a time; the $\gamma$ chain binds to 
the kinase, JAK3, then the $\alpha$ chain binds to the complex formed by $\gamma$ and JAK3. Finally, the ligand, IL-7, binds to the signalling receptor composed of $\gamma$, $\alpha$ and $JAK3$. The model also includes the formation of ``dummy'' receptors, which do not involve the kinase JAK3. Figure~\ref{fig:IL7Rscheme} illustrates the sequential formation of the signalling and dummy complexes. The reaction scheme for this model is as follows
\begin{equation}
    \begin{array}{lllccl}
    \gamma + JAK3 &\rightleftharpoons& \gamma:JAK3,&&&K_1,
    \\
    \alpha + \gamma:JAK3 &\rightleftharpoons& \alpha:\gamma:JAK3,&&&K_2,
    \\
\alpha + \gamma &\rightleftharpoons& \alpha:\gamma,&&&K_2,
\\
    L+\alpha:\gamma&\rightleftharpoons& L:\alpha:\gamma,&&&K_3,
    \\
    L+\alpha:\gamma:JAK3 &\rightleftharpoons& L:\alpha:\gamma:JAK3,&&&K_3,
    \\
\end{array}
\label{net:model1}
\end{equation}
where for $i=1,2,3$, $K_i$ is the affinity constant of the appropriate
reaction.
One can show that this system has deficiency zero and is reversible (see Section~\ref{sec:CRNT}). Therefore, for every set of rate constants and  initial conditions, there exists exactly one positive steady state. Moreover, this positive steady state is in detailed balance.
We remind the reader that in this paper 
we assume mass action kinetics to determine
reaction rates.
We denote the concentration of $\gamma$, $\alpha$, JAK3 and IL-7 by $x$, $y$, $z$, and $L$, respectively. The reaction rate for the forward/backward reaction ($\rightharpoonup$/$\leftharpoondown$) is given by $k_i$ and $q_i$, respectively, for $i=1,2,3$. We note that $K_i = k_i/q_i$. The concentrations of the product complexes of the forward reactions are denoted by $c_i$ in order of appearance (see Figure~\ref{fig:IL7Rscheme}). We can now write down the ordinary differential equations (ODEs) associated to the system of reactions~\eqref{net:model1}:
\begin{equation}
\begin{aligned}
    \frac{dx}{dt}&=-k_1xz+q_1c_1-k_2xy+q_2c_3,
    \\
    \frac{dy}{dt}&=-k_2yc_1+q_2c_2-k_2xy+q_2c_3,
    \\
    \frac{dz}{dt}&=-k_1xz+q_1c_1,
    \\
    \frac{dc_1}{dt}&=k_1xz-q_1c_1-k_2yc_1+q_2c_2,
    \\
    \frac{dc_2}{dt}&=k_2yc_1-q_2c_2-k_3c_2L+q_3c_5,
    \\
    \frac{dc_3}{dt}&=k_2xy-q_2c_3- k_3c_3L+q_3c_4,
    \\
    \frac{dc_4}{dt}&=k_3c_3L-q_3c_4,
    \\
    \frac{dc_5}{dt}&=k_3c_2L-q_3c_5.
    \end{aligned}
    \label{eq:il7r_ODES}
\end{equation}
A suitable basis for the conservation equations is
\begin{equation}
    \begin{aligned}
    N_x&=x+c_1+c_2+c_3+c_4+c_5,
    \\
    N_y&=y+c_2+c_3+c_4+c_5,
    \\
    N_z&=z+c_1+c_2+c_5;
    \end{aligned}
\end{equation}
that is, single chain molecules are conserved since we do not consider the generation or degradation of molecules. The constants $N_x$, $N_y$ and $N_z$ represent the total copy number of $\gamma$, $\alpha$ and JAK3 molecules per cell, respectively.
Detailed balance leads to the following steady state equations:
\begin{equation}
    \begin{aligned}
    c_1&=K_1xz,\\
    c_2&=K_2yc_1,\\
    c_3&=K_2xy,\\
    c_4&=K_3Lc_3,\\
    c_5&=K_3Lc_2.
    \end{aligned}
    \label{eq:il7r_sseq}
\end{equation}
Substituting the steady state equations into the conservation equations, we obtain a system of polynomials
 \begin{equation}
 \begin{aligned}
 0 &= - N_x + x + 
  K_1xz+ K_2K_1xyz +
 K_2xy  + K_3K_2Lxy + K_3K_2K_1Lxyz,
 \\
0 &= - N_y + y +  K_2K_1xyz +
 K_2xy  + K_3K_2Lxy + K_3K_2K_1Lxyz,
 \\
0 &= - N_z + z + K_1xz + K_2K_1xyz + K_3K_2K_1Lxyz.
\end{aligned}
\label{eq:il-7r}
 \end{equation}

\paragraph{Analytic computation
of the steady state.}

The polynomial system \eqref{eq:il-7r} can be solved numerically for a particular set of parameter values. However, an analytic solution
 will provide greater insight and will allow us to derive expressions for the amplitude and the EC$_{50}$. We make use 
of Macaulay2~\cite{M2} to compute a lex Gr\"obner basis for
this model, which will lead to a triangular set of polynomials~\footnote{Example code is provided in Appendix~\ref{appendix:M2code}.}, as follows:
\begin{subequations}
    \begin{align}
     0 &= z^2 + \frac{[1 + K_1 (N_x-N_z)]}{K_1}z - \frac{N_z}{K_1},\label{eq:Mod1_B1}
     \\
     0 &= y^2 + \frac{[1 + K_2( K_3 L+1)(N_x -N_y)]}{K_2(K_3L+1)}y - \frac{N_y}{K_2(K_3L+1)},\label{eq:Mod1_B2}
     \\
     0 &= x - \frac{1}{N_x}yz - \frac{(N_x-N_z)}{N_x}y - \frac{(N_x-N_y)}{N_x}z - \frac{N_x(N_x-N_y-N_z)+N_yN_z}{N_x}.\label{eq:Mod1_B3}
     \end{align}
     \label{eq:MOD1_gb}
 \end{subequations}
Equation~\eqref{eq:Mod1_B3} gives $$x=\frac{(N_x-N_y+y)(N_x-N_z+z)}{N_x} = \frac{N_x-N_y+y}{1+K_1z},$$
where the last equality follows from equation~\eqref{eq:Mod1_B1}. Solving the system~\eqref{eq:MOD1_gb} and selecting the biologically relevant solution, we obtain an analytic expression for the number of
free (unbound) JAK3, $\alpha$ and $\gamma$ molecules at steady state
 \begin{subequations}
  \begin{align}
    z&=\frac{-1 +K_1(N_z-N_x) + \sqrt{\Delta_1}}{2K_1},\\
    y&=\frac{-1+K_2(N_y - N_x)(K_3L + 1) + \sqrt{\Delta_2} }{2K_2(K_3L + 1) },\\
    x&=\frac{N_x-N_y+y}{1+K_1z},
    \intertext{where we have introduced}
    \Delta_1&=4 K_{1} N_z + \left[K_{1} (N_x - N_z) + 1\right]^{2},\nonumber\\
    \intertext{and}
    \Delta_2&=4 K_{2} N_y \left(K_{3} L + 1\right) + \left[K_2(N_x - N_y)(K_3L + 1) + 1\right]^{2}.\nonumber
  \end{align}
  \label{eq:il7rss}
 \end{subequations}
We study the dose-response curve of this model
given by the number of signalling complexes, $L:\gamma:\alpha:JAK$, per cell at steady state and as a function of $L$. The signalling function, $\sigma(L)$, is given by
 \begin{equation}
     \sigma(L)\equiv c_5=K_3K_2K_1Lxyz.
 \end{equation}. 
 
\paragraph{Analytic computation
of the   amplitude.}

A simple inspection of the behaviour of
\eqref{eq:il7rss} shows that the
 dose-response curve is a sigmoid,
 such that
  $\sigma(0) = 0$. Therefore the amplitude
 $A$ is given by the 
 asymptotic behaviour 
of the signalling function as follows:
 \begin{equation}
    A\equiv \lim_{L\to+\infty}\sigma(L).
   \end{equation}
We will prove this result rigorously for a more general class of models in Section~\ref{sec:generalisation-SRLK}.

We first notice that $z$ is independent of $L$. We now compute the product $xy$ (at steady state) as follows
 \[xy=\frac{(N_x-N_y)y+y^2}{1+K_1z}.\]
From equation~\eqref{eq:Mod1_B2} we can replace the polynomial in $y$ of degree two by an expression linear in $y$:
\[(N_x-N_y)y+y^2=\frac{N_y-y}{K_2(K_3L+1)}.\]
Thus, we obtain the following analytic expression for the 
signalling function:
 \begin{equation}
     \sigma(L)=K_3K_2K_1Lxyz=\frac{K_1z}{(1+K_1z)}\frac{K_3L}{(K_3L+1)}(N_y-y).
     \label{eq:il7rsig}
 \end{equation}
Since $\frac{K_3L}{1+K_3L}\to 1$ when $L\to+\infty$, we need to study the expression $N_y-y$ in this limit. We have
\begin{equation}
N_y-y=\frac{(N_y+N_x)K_2(K_3L+1)+1-\sqrt{\Delta_2}}{2K_2(K_3L+1)},
\label{eq:il7r_ny-y}
\end{equation}
where \[\Delta_2=K_2^2(K_3L+1)^2(N_x-N_y)^2+2K_2(K_3L+1)(N_x+N_y)+1.\]
Keeping  to lowest order in $\mathcal{O}(\frac{1}{L})$ we obtain
\begin{align}
N_y-y&=\frac{1+(N_x+N_y)K_2(K_3L+1)-K_2(K_3L+1)|N_x-N_y|(1+\mathcal{O}(\frac{1}{L}))}{2K_2(K_3L+1)},\\
&=\frac{N_x+N_y-|N_x-N_y|}{2}+\mathcal{O}(\frac{1}{L}).
\end{align}
Finally, noticing that \[\frac{N_x+N_y-|N_x-N_y|}{2}=\text{min}(N_x,N_y),\] 
we obtain the amplitude
\begin{equation}
    A=\text{min}(N_x,N_y)\frac{K_1z}{1+K_1z},
\end{equation}
where $z$ is the analytic expression obtained in \eqref{eq:il7rss}.
This result indicates that the amplitude of this model is the total number of the limiting trans-membrane chain modulated by a factor,
valued in the interval $[0,1]$, which only depends on $K_1$, $N_x$ and $N_z$. 

\paragraph{Analytic computation
of the  
 EC$_{50}$.}

We now determine the EC$_{50}$ by finding the value of $L_{50}$ such that 
\begin{equation}
\sigma(L_{50})=\frac{A}{2} = K_1K_2K_3L_{50}x_{50}y_{50}z_{50},
  \label{eq:il7r_ec50def}
\end{equation}
where $x_{50}$, $y_{50}$ and $z_{50}$ are the steady state expressions found in \eqref{eq:il7rss} evaluated at $L=L_{50}$.
Two expressions satisfy this equation but only one provides
a relevant biological  solution with $L,x,y,z > 0$.
The relevant analytic expression of the EC$_{50}$ is given by
\begin{equation}
     EC_{50}=M\; 
     \frac{1+K_2(N_x+N_y-M)+\sqrt{1+K_2^2(N_y-N_x)^2+2K_2(N_x+N_y-M)}}{K_2K_3(M-2N_x)(M-2N_y)},
     \label{eq:il7r_ec50}
\end{equation}
with $M=\text{min}(N_x,N_y)$. The details of the computation can be found in Appendix~\ref{appendix:il7r}. 
This result shows that 
 the EC$_{50}$ value for this system is independent of the kinase,
 since the parameters $K_1$ and $N_z$ are absent in the previous
 expression.

Alternatively, we now propose a more algebraic method 
to derive the analytic expression of the EC$_{50}$. We compute a lex Gr{\"o}bner basis for the augmented system of polynomials consisting of the steady state equations~\eqref{eq:il-7r} and
\begin{equation}
       K_1K_2K_3Lxyz \; (1+K_1z)-\frac{MK_1z}{2}  = 0,
\end{equation}
where this time $x$, $y$, $z$, and $L$ are variables. The resulting triangular system describes directly the EC$_{50}$ and $x$, $y$, $z$ at $L=$EC$_{50}$:
\begin{subequations}
    \begin{align}
        0&=L^2+\frac{2M[-1+K_2(M-N_x-N_y)]}{K_2K_3(M-2N_x)(M-2N_y)}L+\frac{M^2}{K_3^2(M-2N_x)(M-2N_y)}, \label{eq:Mod1_gb_EC50L}\\
        0&=z^{2}+\frac{1+K_1(N_x-N_z)}{K_1}z-\frac{N_z}{K_1} ,\label{eq:Mod1_gb_EC50z}\\
        0&=y-\frac{K_3(M-2N_x)(M-2N_y)}{2M}L+\frac{K_2(2N_x-M)+2}{2K_2}, \label{eq:Mod1_gb_EC50y}\\
        0&=x-\frac{K_3(M-2N_x)(M-2N_y)(N_x-N_z+z)}{2MN_x}L+\frac{[2+K_2(2N_y-M)](N_x-N_z+z)}{2K_2N_x}.\label{eq:Mod1_gb_EC50x}
    \end{align}
\end{subequations}
Solving \eqref{eq:Mod1_gb_EC50L} and selecting the solution for which $y$ and $x$, given by equations \eqref{eq:Mod1_gb_EC50y} and \eqref{eq:Mod1_gb_EC50x}, respectively, are positive yields the final result,
in agreement with \eqref{eq:il7r_ec50}.
 
 
\subsubsection{The  IL-7 receptor-ligand system: an additional  sub-unit receptor chain}
\label{sec:il7r+r}

\begin{figure}[htp!]
    \centering
          \begin{subfigure}{.3\textwidth}
    \includegraphics[width=\textwidth]{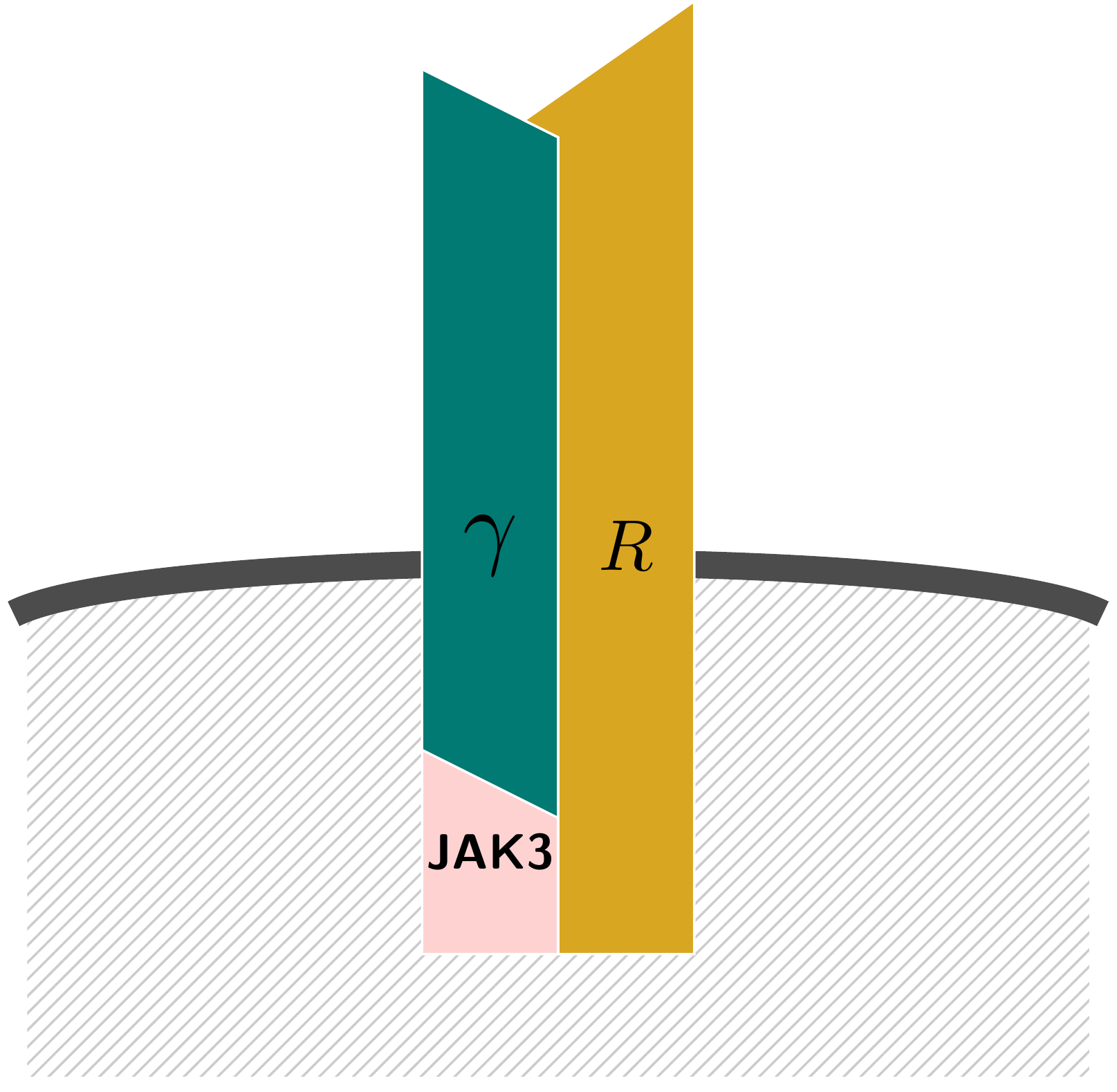}
    \caption{Decoy complex with kinase.}
    \label{fig:IL7rrdecoyk}
  \end{subfigure}
 \begin{subfigure}{.3\textwidth}
    \includegraphics[width=\textwidth]{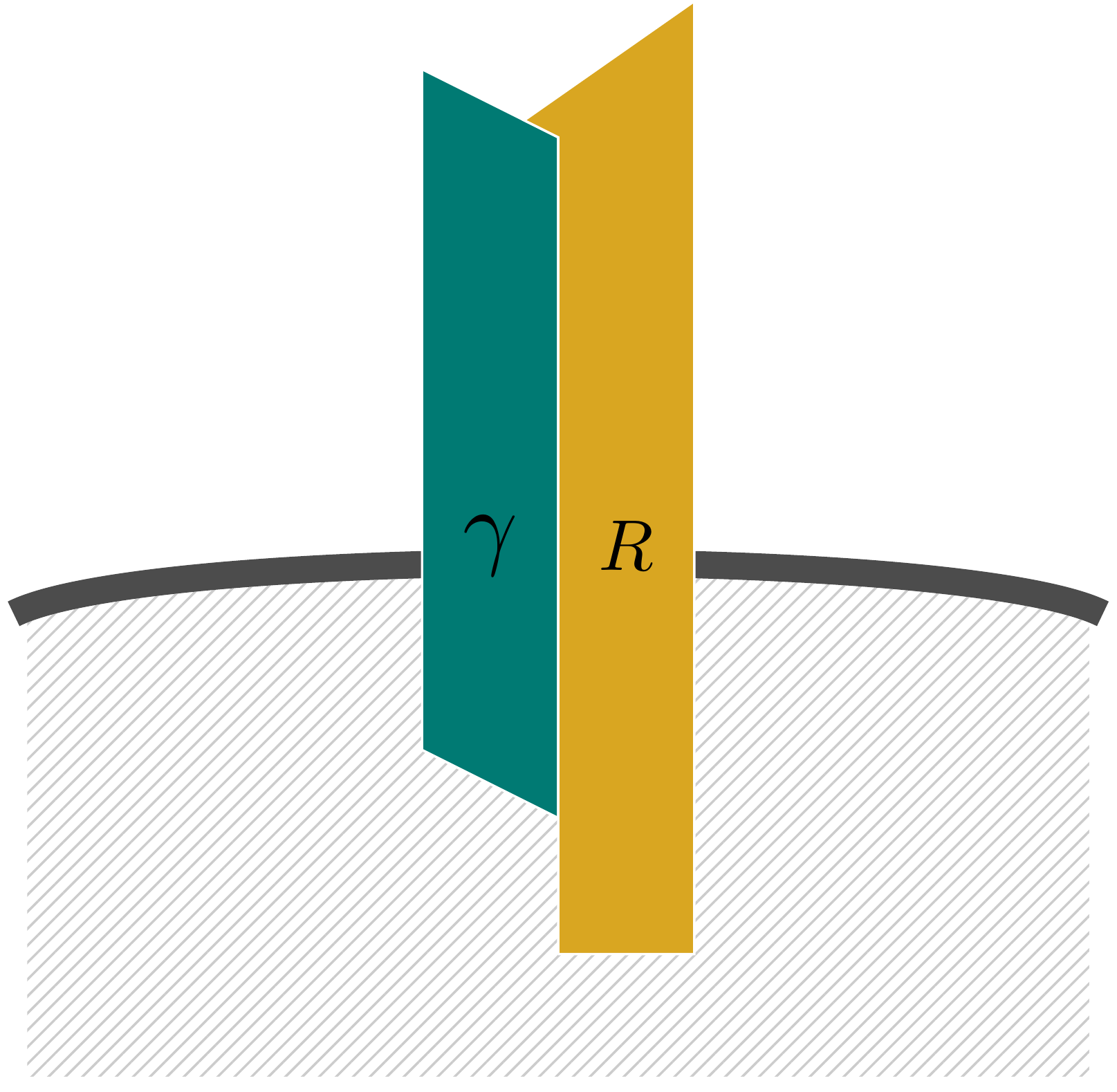}
    \caption{Decoy complex without kinase.}
    \label{fig:IL7rrdecoynok}
  \end{subfigure}
   \begin{subfigure}{\textwidth}
    \includegraphics[width=\textwidth]{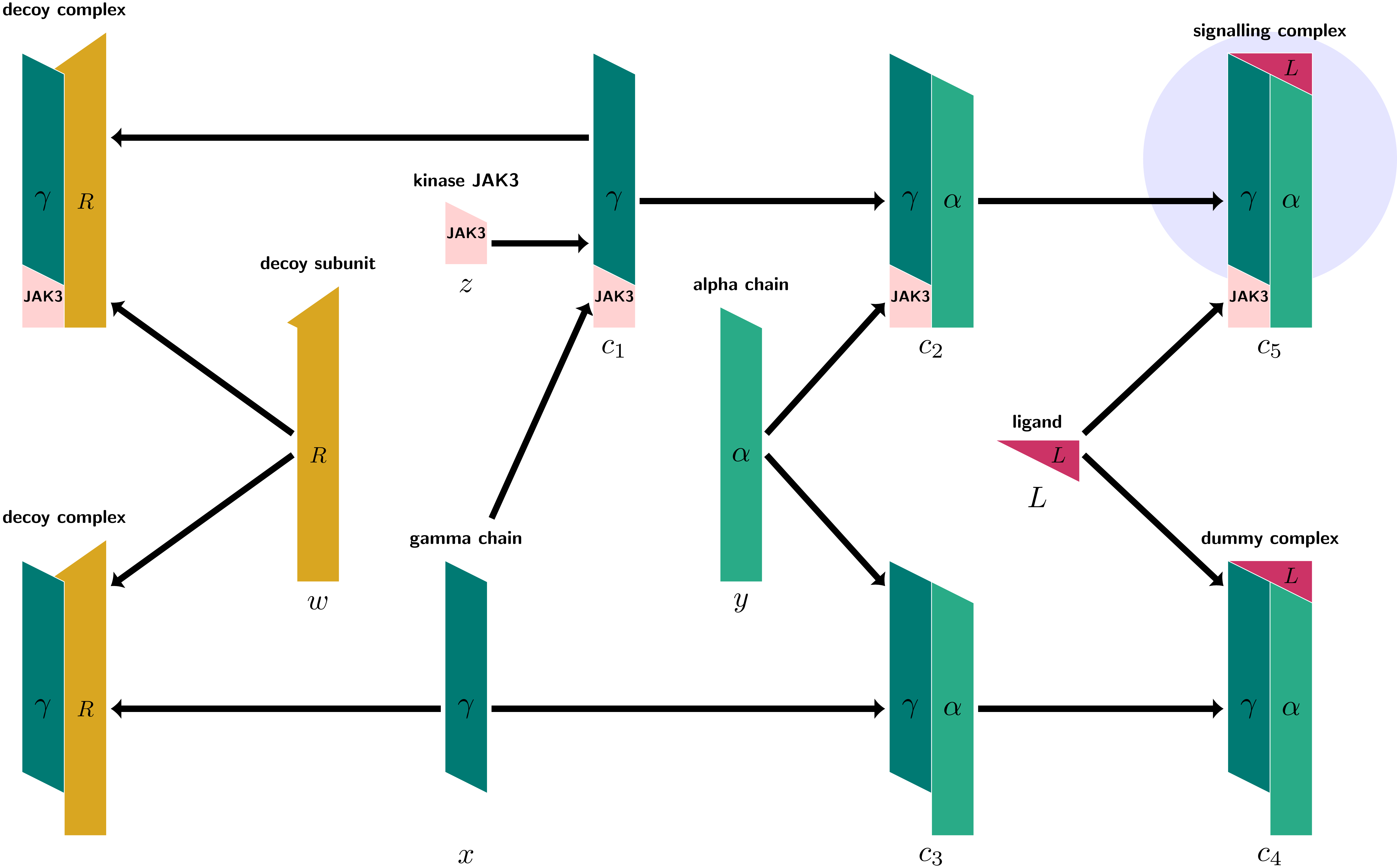}
    \caption{Second IL-7R model: sequential chemical reaction scheme.}
    \label{fig:IL7rrscheme}
  \end{subfigure}
    \caption{IL-7R model with an additional receptor sub-unit. The signalling and dummy complexes are the same as in the first IL-7R model. This second model allows the formation of decoy complexes: (a) with the kinase JAK3,  or (b) without the kinase. (c) The IL-7R dummy and signalling complexes are formed sequentially. Decoy complexes can be formed to prevent the formation of signalling or dummy complexes. The mathematical notation used is annotated below each molecule or complex.}
    \label{fig:IL7R+R}
\end{figure}

The previous model described the IL-7 receptor system without any consideration for the fact that the $\gamma$ chain is shared with other cytokine receptors~\cite{rochman2009new}. We now account for this competition by including in the previous model an additional receptor chain, $R$, which can bind to the $\gamma$ chain, or the complex $\gamma:JAK3$, to form decoy receptor complexes (see Figure~\ref{fig:IL7rrdecoyk} and Figure~\ref{fig:IL7rrdecoynok}, where the hatched area indicates the cytoplasmic region).
The resulting reaction scheme  (summarised in Figure~\ref{fig:IL7rrscheme}) is given by
\begin{equation*}
    \begin{array}{lllccl}
    JAK3 + \gamma &\rightleftharpoons& JAK3:\gamma&&&K_1,
    \\
    \alpha + JAK3:\gamma &\rightleftharpoons& \alpha:\gamma:JAK3&&&K_2,
    \\
    \alpha + \gamma &\rightleftharpoons& \alpha:\gamma&&&K_2,
    \\
    L+\alpha:\gamma&\rightleftharpoons& L:\alpha:\gamma&&&K_3,
    \\
    L+\alpha:\gamma:JAK3 &\rightleftharpoons& L:\alpha:\gamma:JAK3&&&K_3,\\
    R+\gamma&\rightleftharpoons& R:\gamma &&& K_4,
    \\
    R+JAK3:\gamma &\rightleftharpoons& R:\gamma:JAK3 &&& K_4.
\end{array}
\end{equation*}
We use $w$ to describe the concentration of the additional chain $R$. Similarly to the previous model, we write the  system
of ODEs describing the time evolution for each complex  and then derive (a basis for) the conservation and steady state equations. Combining them, we obtain the following polynomial system: 
\begin{equation}
\begin{aligned}
0 &= - N_x + x + K_2xy + K_1xz + K_2K_1xyz + K_3K_2Lxy + K_3K_2K_1Lxyz+K_4xw+K_1K_4xwz,\\
0 &= - N_y + y + K_2xy + K_2K_1xyz + K_3K_2Lxy + K_3K_2K_1Lxyz,\\
0 &= - N_z + z + K_1xz + K_2K_1xyz + K_3K_2K_1Lxyz+K_1K_4xwz,\\
0 &= -N_w+w+K_4xw+K_4K_1xwz,
 \end{aligned}
 \label{eq:il7r+rpoly}
\end{equation}
where $N_w$ is the additional conserved quantity.
Again, we compute a lex Gr\"obner basis for this set of polynomials to obtain the following triangular system:
\begin{subequations}
\begin{align}
0 &= K_1z^2 + z [1+ K_1 (N_x-N_z)] -N_z,\label{eq:P1}
\\
0 &= Ay^3+By^2+Cy+D,\label{eq:P2}
\\
0 &= \left[K_2K_4(1+K_3L)N_xN_y\right]x+(Ay^2+By+C+K_4N_y)(N_x-N_z+z),\label{eq:P3}\\
0 &= \left[K_2K_4(1+K_3L)N_y\right]w+Ay^2+By+[K_2(1+K_3L)-K_4]N_y,\label{eq:P4}
 \end{align}
 \label{eq:il7+r-GB}
 \end{subequations}
 where
 \begin{align*}
     A&=-K_2 (1 + K_3 L) [K_2(1+K_3L) - K_4],\\
     B&=K_4 - K_2 (1 + K_3 L) [1 + K_4 (N_w - N_x + 2 N_y) +K_2 (1 + K_3 L) (N_x - N_y)],\\
     C&=N_y [-2 K_4 + K_2 (1 + K_3 L) (1 + K_4 (N_w - N_x + N_y))],\\
     D&= K_4 N_y^2.
 \end{align*}
Solving~\eqref{eq:P1} gives the number of free JAK3 molecules per cell at steady state, $z$; solving~\eqref{eq:P2} gives the number of free  (unbound) $\alpha$ chains per cell, $y$;
and substituting $y$ and $z$ into \eqref{eq:P3} and \eqref{eq:P4} gives the remaining steady states.
We obtain the following implicit steady state expressions for the number of free (unbound) chains: 
\begin{equation}
\begin{aligned}
    z=&\frac{-1+K_1(N_z-N_x)+\sqrt{[1+K_1(N_x-N_z)]^2+4N_zK_1}}{2K_1},
    \\
    x=&-\frac{(Ay^2+By+C+K_4N_y)(N_x-N_z+z)}{K_2K_4(1+K_3L)N_xN_y},
    \\
    w=&-\frac{Ay^2+By+[K_2(1+K_3L)-K_4]N_y}{K_2K_4(1+K_3L)N_y}.
\end{aligned}
\label{eq:il7r+rss}
\end{equation}
The problem now reduces to finding the positive real roots of~\eqref{eq:P2}. As~\eqref{eq:P2} is a polynomial of degree 
three, we could, in principle, find an exact analytic solution. However, such a solution might not be very informative. Instead, we show how perturbation theory can be used to obtain the amplitude of the dose-response.
In this model, the signalling complex is still  $L:\alpha:\gamma:JAK3$. The signalling function is given by
\begin{equation}
    \sigma(L)\equiv K_3K_2K_1Lxyz.
\end{equation}
In Section~\ref{sec:SRLK-extensions} we will show that, for this model, the maximum of $\sigma$ is attained in the limit $L\to+\infty$. Hence, we have
\begin{equation}
    A\equiv \lim_{L\to+\infty}\sigma(L).
\end{equation}
Combining \eqref{eq:P1}, written as $N_x-N_z+z=\frac{N_x}{1+K_1z}$, and \eqref{eq:P2}, we obtain a reduced expression for the product $xy$
\begin{equation}
    xy=\frac{N_y-y}{K_2(1+K_3L)(1+K_1z)},
\end{equation}
which allows us to rewrite the amplitude  as
\begin{equation}
    A=\lim_{L\to+\infty}\frac{K_1z}{(1+K_1z)}\frac{K_3L}{(K_3L+1)}(N_y-y).
\end{equation}
We note that $z$ is independent of $L$ and, therefore, to compute the amplitude we only need to find the behaviour of $y$ as $L\to+\infty$.

\paragraph{Perturbation theory to determine $y$ as $L\to +\infty$.} 

\begin{figure}[htp!]
    \centering
    \includegraphics[width=0.5\textwidth]{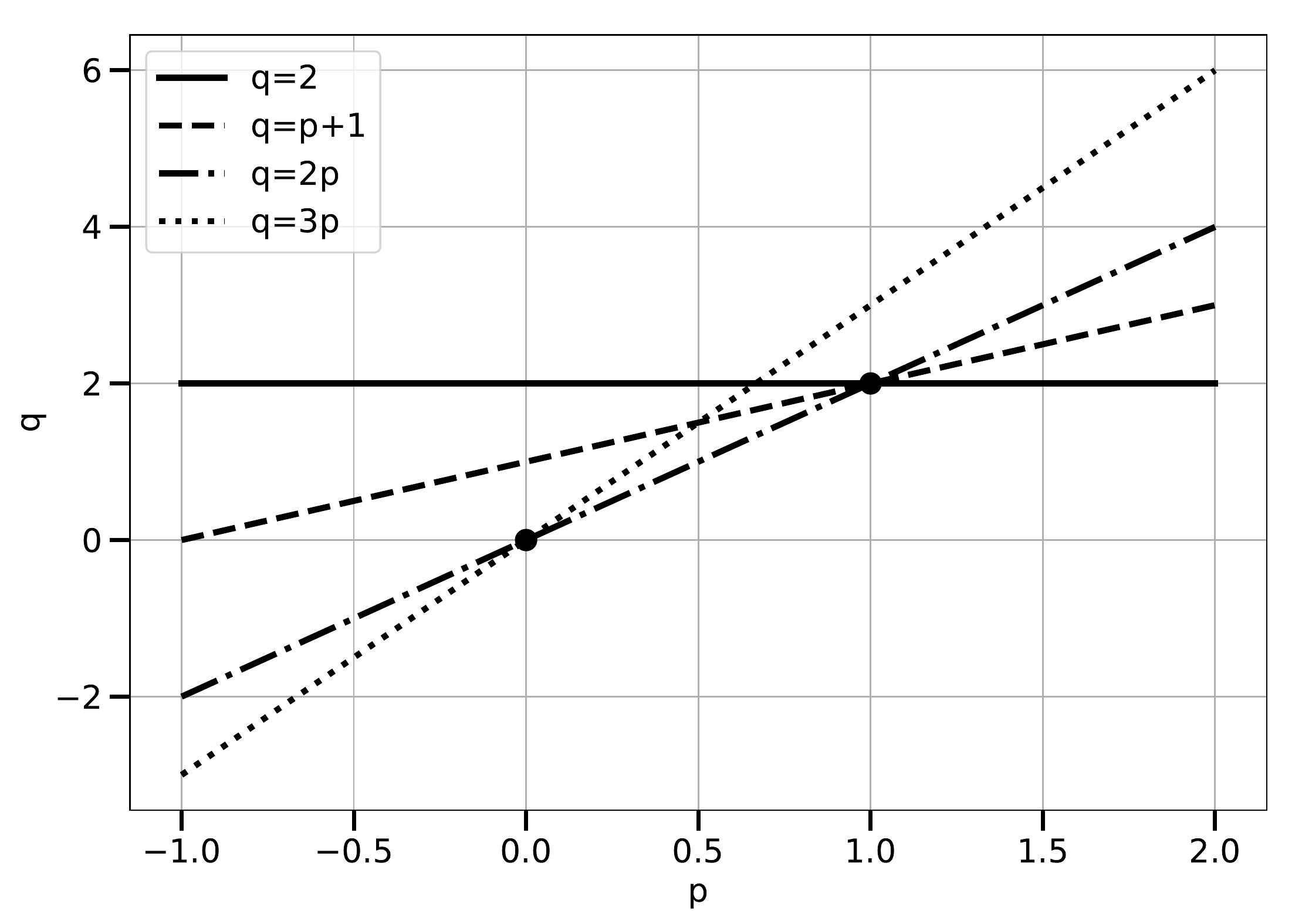}
    \caption{The lines defined in set $E$ and the proper values, (black dots), computed following the graphical algorithm described in Appendix~\ref{appendix:perturbationtheory}.}
    \label{fig:propervalues_modelwithK6}
\end{figure}

We now apply the method described in  Ref.~\cite{simmonds2013first} and summarised in Appendix~\ref{appendix:perturbationtheory}. Let $\epsilon=\tfrac{1}{L}$ and define the polynomial $P_\epsilon$ as follows:
\[P_\epsilon(y)\equiv P_2(y)\epsilon^2,\] 
where $P_2$ is the polynomial \eqref{eq:P2}. We added a factor of $\epsilon^2$ to remove any negative powers of $\epsilon$ in $P_2$.  We obtain the polynomial
 \begin{equation}
     P_\epsilon(y)=A_\epsilon y^3+B_\epsilon y^2+C_\epsilon y+D_\epsilon,
 \end{equation}
where
\begin{align*}
     A_\epsilon=&  -K_2 (\epsilon +   K_4) [K_2 (\epsilon + K_4) - \epsilon K_6],
     \\
     B_\epsilon=& K_2^2 K_4^2 (N_y-N_x) - 
  \epsilon K_2 K_4 [1 + 2 K_2 N_x - 2 K_2 N_y + K6 (N_w - N_x + 2 N_y)] \\+ 
  &\epsilon^2 (K_6 - K_2 (1 + K_2 N_x - K_2 N_y + K_6 (N_w - N_x + 2 N_y))),\\
     C_\epsilon=& \epsilon N_y (K_2 K_4 (1 + K_6 (N_w - N_x + N_y)) + 
    \epsilon (K_2 - 2 K_6 + K_2 K_6 (N_w - N_x + N_y))),\\
     D_\epsilon=& K_6 N_y^2 \epsilon^2.
 \end{align*}
We now replace $y$ by $\epsilon^p \omega(\epsilon)$ with $\omega(0)\neq 0$ according to theorem~\ref{th:theoremroot}.
We obtain 
\begin{equation}
    P_\epsilon(\epsilon^p\omega(\epsilon))=A_{p,\epsilon} \omega^3+B_{p,\epsilon} \omega^2+C_{p,\epsilon} \omega+D_{p,\epsilon},
\end{equation}
where
\begin{align*}
     A_{p,\epsilon}=& -\epsilon^{3 p} K_2 (\epsilon + K_4) (K_2 (\epsilon + K_4) - \epsilon K_6),
     \\
     B_{p,\epsilon}=& \epsilon^{2 p} (K_2^2 K_4^2 (N_y-N_x) - 
   \epsilon K_2 K_4 (1 + 2 K_2 (N_x-N_y) + K_6 (N_w - N_x + 2 N_y))
  \\
   +& \epsilon^2 (K_6 - K_2 (1 + K_2 N_x - K_2 N_y + K_6 (N_w - N_x + 2 N_y)))),
   \\
     C_{p,\epsilon}=& \epsilon^{1 + p} N_y (K_2 K_4 (1 + K_6 (N_w - N_x + N_y)) + 
   \epsilon (K_2 - 2 K_6 + K_2 K_6 (N_w - N_x + N_y))),
   \\
     D_{p,\epsilon}=& K_6 N_y^2 \epsilon^2.
 \end{align*}
The smallest exponents in the previous equation
are \[E=\{2,1+p,2p,3p\}.\] 
We note that $0$ is not in $E$ because we multiplied $P_2$ by $\epsilon^2$. Applying the graphical algorithm detailed in Appendix~\ref{appendix:perturbationtheory}, we find the proper values $(0,0)$ and $(1,2)$ (see Figure~\ref{fig:propervalues_modelwithK6}). We investigate these two branches. 

\paragraph{Branch (0,0).}
We make use of  the notation in Appendix~\ref{appendix:perturbationtheory}, to define
\[T_\epsilon^{(1)}(\omega) \equiv \epsilon^0 P_\epsilon(\omega\epsilon^0).\] The least common denominator of \{2,1,0,0\} is
$q_1=1$. Therefore in accordance with the notation of Appendix~\ref{appendix:perturbationtheory}
\[\epsilon=\beta,\] and
the polynomial $R_\beta^{(1)}$ defined as
\[ R_\beta^{(1)}(\omega)\equiv T_\epsilon^{(1)}(\omega),\] 
is the polynomial $P_\epsilon$. It means that we have $y=\omega$ and we can directly carry out a regular perturbation expansion.

Let us write the asymptotic expansion $y=y_0+y_1\epsilon+y_2\epsilon^2+\ldots $ and replace it in $P_\epsilon(y)$.  Since $P_\epsilon(y)=0$, by the fundamental theorem of perturbation theory (Theorem~\ref{th:fundamentalTheoremPerturbation}) we obtain a system of equations in $y_0$, $y_1, \ldots $, which can be solved. The first equation of the system is given by
\begin{equation}
    -K_2^2 K_3^2 y_0^2 (N_x - N_y + y_0)=0.
    \label{eq:y0_branch1}
\end{equation}
We are are only interested in non-negative  values of $y_0$, since we want $y$ to be biologically relevant. We also require $\omega(0)=y(0)=y_0\neq 0$ from Theorem~\ref{th:theoremroot}. Thus, solving equation~\eqref{eq:y0_branch1}, we obtain $y_0=N_y-N_x$ if $N_y>N_x$ and $y_0=0$ otherwise. 
Assuming $y_0=0$ ({\em i.e.,} $N_x\geq N_y$), we solve the next order equation
\begin{equation}
 K_4N_y^2+K_2K_3N_y(1+K_4(N_w - N_x + N_y))y_1+K_2^2K_3^2(N_y-N_x)y_1^2=0.
 \label{eq:y1_branch1_y0=0}
\end{equation}
We select the positive root of this polynomial and obtain an expression for $y_1$ when $y_0=0$. Thus, we have
\begin{equation}
    y_1=N_y\frac{1 + K_4 (N_w - N_x + N_y) + \sqrt{1 + 2 K_4 (N_w + N_x - N_y) + K_4^2 (N_w - N_x + N_y)^2}}{2K_2K_3(N_x-N_y)}.
    \label{eq:b1_y1}
\end{equation}
Equation \eqref{eq:b1_y1} shows that $y_1 > 0$ when $N_x \geq N_y$. Hence, $y \approx \epsilon y_1$ converges to zero from above and, therefore, represents a biologically relevant solution.
We can conclude that
\begin{equation}
\lim_{\epsilon\to0}(N_y-y)= \left\{
\begin{array}{ll}
 N_x,& \text{if $N_y>N_x$,}\\
 N_y,& \text{otherwise.} 
\end{array}
\right\} = \min(N_x,N_y).
\end{equation}

\paragraph{Branch (1,2).}

On this branch, and again following the notation of Appendix~\ref{eq:preparedPolynomialPerturbation}, we define \[T_\epsilon^{(2)}(\omega) \equiv \epsilon^{-2} P_\epsilon(\epsilon^1\omega).\] The least common denominator of \{2,1+1,0+1,0+1\} is $q_2=1$, so $R_{\beta}^{(2)}$ is the same polynomial as $T^{(2)}_\epsilon$. Since $y=\omega\epsilon$, we have $N_y-y \isEquivTo{\epsilon \to 0}N_y$.
Furthermore, when replacing $\omega$ by an asymptotic expansion $\omega_0+\omega_1\epsilon+\ldots $ in $T_\epsilon^{(2)}$ and applying the fundamental theorem of perturbation theory (Theorem~\ref{th:fundamentalTheoremPerturbation}), we obtain the same equation for $w_0$ as for $y_1$ in the previous branch (see equation~\eqref{eq:y1_branch1_y0=0}):
\begin{equation}
 K_4 N_y^2 +K_2 K_3N_y(1+K_4(N_w - N_x + N_y))\omega_0+K_2^2K_3^2(N_y-N_x)\omega_0^2=0
\; .
\end{equation}
We have $y_1=\omega_0$. In other words, at large, but finite $L=1/\epsilon$, the convergence behaviour of the two branches is identical. This agrees with Theorem~\ref{thm:DefZero} which states that there is only one positive solution for each set of reaction constants and initial conditions.
In conclusion, we find that $N_y-y=\text{min}(N_x,N_y)$, which gives the following expression for the amplitude 
\begin{equation}
    A\equiv \frac{K_1z}{1+K_1z}\text{min}(N_x,N_y),
\end{equation}
with $z$ defined in \eqref{eq:il7r+rss}. As the steady state concentration of JAK3, $z$, is the same in the IL-7R model with or without the extra chain $R$, the amplitude of both models has the exact same expression.

\paragraph{Computation of the EC$_{50}$.}

Since we did not compute analytic expressions for each
steady state concentration, the EC$_{50}$ expression has to be obtained by computing a Gr\"obner basis of the polynomial system~\eqref{eq:il7r+rpoly} augmented by the polynomial
\begin{equation}
    K_3K_2K_1Lxyz(1+K_1z)-\frac{K_1zM}{2}=0,
\end{equation}
considering $x,y,z$ and $L$ as variables,
with $M=\text{min}(N_x,N_y)$. 
The lex Gr\"obner basis obtained for this system is:
\begin{subequations}
\begin{align}
0&=K_1z^2+(1+K_1(N_x-N_z))z-N_z,\label{eq:polyz}\\
0&=K_3aL^3+A_LL^2+B_LL+C_L, \label{eq:polyL}\\
0&=y+\frac{-aL^2+B_yL+C_y }{2K_2(K_2-K_4)M^2},\label{eq:polyy}\\
0&=w+\frac{aL^2+B_wL+C_w}{2K_4(K_2-K_4)M^2}, \label{eq:polyw}\\
0&=x+\frac{aL^2+B_xL+C_x}{2K_2K_4M^2(1+K_1z)}, \label{eq:polyx}
\end{align}
\end{subequations}
where we wrote:
\begin{align*}
    a&=K_2^2K_3^2(M-2N_x)(M-2N_y)^2,\\
    A_L&=K_2K_3^2M(M-2N_y)(-2 + 3 K_2 M - K_4 (M + 2 N_w - 2 N_x) -  2 K_2 (2 N_x + N_y)),\\
    B_L&=K_3 M^2(2 K_4 + K_2 (-2 + 3 K_2 M + 2 K_4 (-M - N_w + N_x + N_y) - 2 K_2 (N_x + 2 N_y))),\\
    C_L&=K_2(K_2-K_4)M^3,\\
    B_y&=-\frac{A_L}{K_3},\\
C_y&=M^2(-2K_4+K_2(2+K_4(M+2N_w-2N_x)-2K_2(M-N_x-N_y))),\\
B_w&=-2K_2K_3M(M-2N_y)(1+K_4N_w+K_2(N_x+N_y-M)),\\
C_w&=K_2M^2(K_2(M-2N_y)-2K_4N_w),\\
B_x&=-K_2K_3M(M-2N_y)(2+K_4(M+2N_w-2N_x)-2K_2(M-N_x-N_y)),\\
C_x&=M^2(2K_4+K_2(K_2-K_4)(M-2N_y)).
\end{align*}
The polynomial~\eqref{eq:polyz} is expected to  be independent of the ligand concentration, $L$. The EC$_{50}$ expression is the real positive root of polynomial~\eqref{eq:polyL} at which $x$, $y$ and $w$  (obtained via polynomials~\eqref{eq:polyx}, \eqref{eq:polyy} and \eqref{eq:polyw}, respectively) are positive. The polynomial~\eqref{eq:polyL} reflects the parameter dependency of the EC$_{50}$: since the parameters $K_1$ and $N_z$ are not present in its coefficients, we can affirm that the EC$_{50}$ is, once again, independent of the kinase.
Thus, we reduced the problem of computing the EC$_{50}$ to solving a univariate polynomial (equation~\eqref{eq:polyL}). In comparison, before any algebraic manipulation was possible, the polynomial system~\eqref{eq:il7r+rpoly} had to be solved multiple times to obtain the dose-response curve (a sigmoid), which was then fitted with a Hill equation. Finally, the EC$_{50}$ was computed from the fitted parameters of the Hill curve. Alternatively, if one wanted to apply the Gr{\"o}bner basis-free method of Section~\ref{sec:il7r}, one would have to solve the polynomial~\eqref{eq:P2} in $y$ (which is possible in theory), find its positive real solution (which is potentially hard), substitute the expression of $y$ into $\sigma(L)$, and then solve for the EC$_{50}$.

\subsection{Summary of proposed algebraic method to study the signalling function}
\label{sec:general-method}

From the two previous examples, we devise a general algorithm to compute analytic expressions of the steady state, 
the amplitude and the EC$_{50}$ for some receptor-ligand systems when  ligand is in excess.

  \begin{center}
 \fbox{\parbox{0.97\textwidth}{%
\noindent \textbf{\large Key steps}
\vspace{0.3cm}
\begin{enumerate}[leftmargin=30pt, rightmargin=30pt,noitemsep,nolistsep, label={\arabic*)}]
\item Write the mass action kinetics set of ODEs for the system under consideration.

\item  Obtain the polynomial system by combining the steady state and conservation equations.

\item  Compute a lex Gr\"obner basis of the polynomial system obtained in step 2.

\item  Define the signalling function $\sigma(L)$.

\item  Compute the amplitude expression by finding the extreme values of $\sigma$: 
  $$\text{Amplitude}=\max(\sigma)-\min(\sigma).$$
  
\item  Compute a lex Gr\"obner basis of the polynomial system obtained in step 2 augmented by the equation $$\sigma(L)- \left[ \frac{\text{Amplitude}}{2} +\min(\sigma) \right] = 0,$$
 with the ligand concentration, $L$, considered an additional variable. This additional equation corresponds to definition~\ref{def:ec50} of the EC$_{50}$.
  
\item  Find the positive roots of the univariate polynomial in $L$ of the Gr\"obner basis obtained in step 6. The root which allows the other variables  to be positive is the EC$_{50}$. 
\end{enumerate}}}
\end{center}

One of the crucial parts of the proposed algebraic algorithm is the amplitude computation. Usually, we have the simplification that $\min(\sigma)=\sigma(0)=0$, however, finding $\max(\sigma)$ can be challenging. For certain classes of models we have
\[\lim_{L\to+\infty}\sigma(L)=\text{max}(\sigma),\]
which greatly reduces the calculation. We can now either solve the Gr\"obner basis from step 3 directly, to obtain  analytic expressions of the steady state concentrations of the single chains components, or use perturbation theory as outlined in Section~\ref{sec:il7r+r}.
In  the final step, if an exact expression for the EC$_{50}$ cannot be computed, {\em i.e.,} the univariate polynomial in $L$ has a large degree, one already reduces the cost of the EC$_{50}$ computation compared to the naive approach.
In summary,
in this section
we compute the lex Gr\"obner bases with the computer algebra package Macaulay2~\cite{M2} and provide a Macaulay2 code example in Appendix~\ref{appendix:M2code}.

\section{Analytical study of general sequential receptor-ligand systems}
\label{sec:generalisation-SRLK}

In spite of the general applicability of the method outlined in the previous section, we still have to make the assumption that the computed limit of the signalling function coincides with its amplitude.
In this section we show that this is indeed the case for a wider class of receptor-ligand systems. An analytic closed-form expression for the amplitude follows with little extra work. The EC$_{50}$ can then be studied
making use of the extended Gr{\"o}bner basis introduced in Section~\ref{sec:general-method}.
We start by giving an abstract generalisation of the example from Section~\ref{sec:il7r}.

\begin{definition}[SRLK model]
We call a \emph{sequential receptor-ligand model with extrinsic kinase (SRLK)} a receptor-ligand model with the following properties:
\begin{itemize}

    \item The receptor is composed of $n$ different trans-membrane chains, $X_1,\ldots ,X_n$, which bind sequentially, $$X_1:\ldots :X_{i-1}+X_i\rightleftharpoons X_1:\ldots :X_i \ \ \text{ for all } i\in\{2,\ldots ,n\}.$$
    
    \item $X_1$ can bind reversibly to an intra-cellular extrinsic kinase $Z$.
    
    \item The  signalling receptor is given by $Z:X_1:\ldots :X_n$ and the  dummy receptor by $X_1:\ldots :X_n$.
    
    \item The extra-cellular ligand, $L$, binds reversibly to the
     signalling (or dummy) receptor, forming the signalling (or dummy) complex $Z:X_1: \ldots :X_n:L$ (or $X_1:\ldots :X_n:L$).
\end{itemize}
\label{def:SRLK}
\end{definition}

\begin{figure}[htp!]
         \centering
         \begin{subfigure}[b]{\textwidth}
         \includegraphics[width=\textwidth]{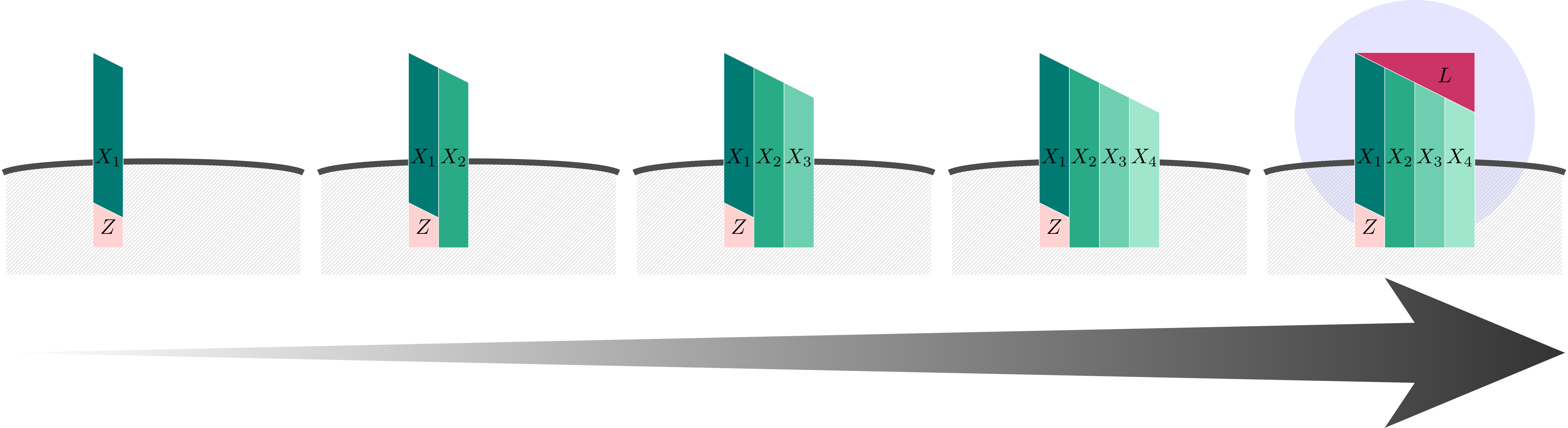}
         \caption{Sequential formation of the signalling complex.}
         \label{fig:SRLKsig}
         \end{subfigure}
                  \begin{subfigure}[b]{\textwidth}
         \includegraphics[width=\textwidth]{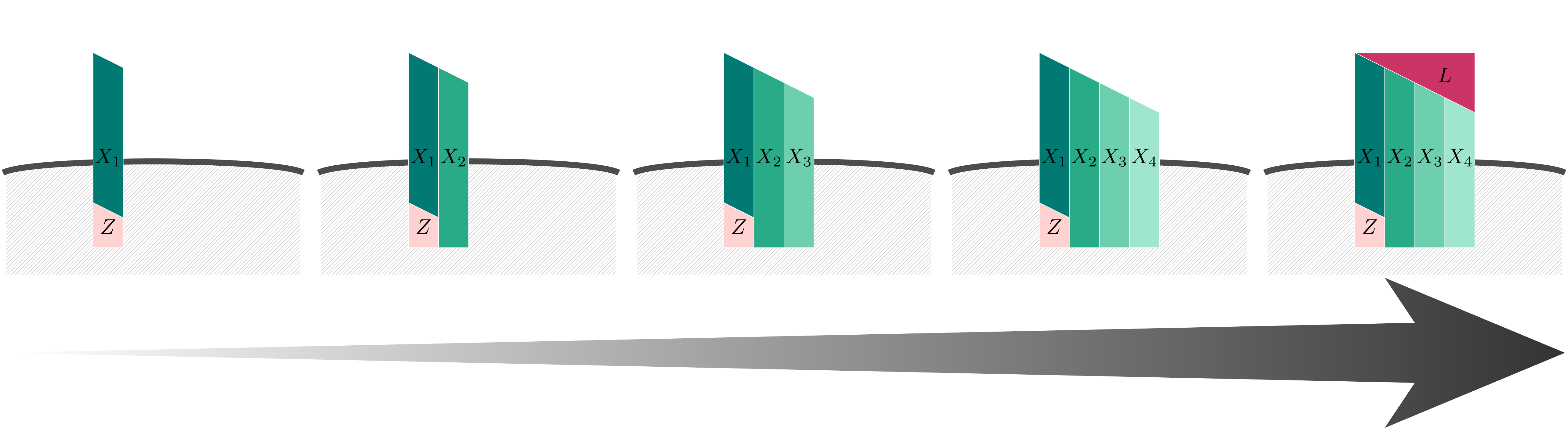}
         \caption{ Sequential formation of the dummy complex.}
         \label{fig:SRLKdum}
         \end{subfigure}
                  \begin{subfigure}[b]{\textwidth}
         \includegraphics[width=\textwidth]{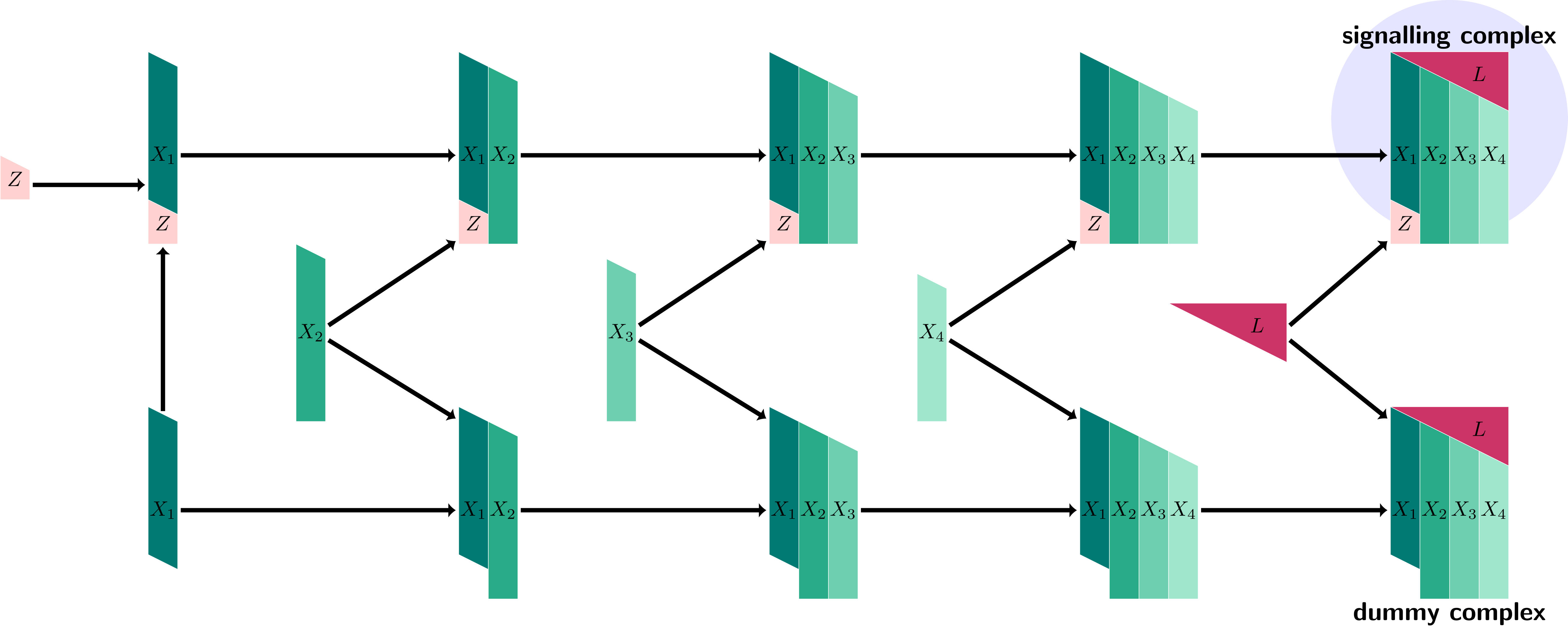}
         \caption{SRLK model: sequential biochemical reaction scheme.}
         \label{fig:SRLKreac}
         \end{subfigure}
  \caption{SRLK model with $n=4$ trans-membrane chains: (a)/(b) Sequential formation of the signalling/dummy complex. (c) Scheme of the sequential formation of the signalling/dummy complex: the chain $X_1$ binds first to the intra-cellular extrinsic kinase $Z$ (only for the signalling complex). Next, the chain $X_2$ binds to the complex $Z:X_1$ (or $X_1$) and  $X_3$ binds to $Z:X_1:X_2$ (or $X_1:X_2$). Then, $X_4$ binds to $Z:X_1:X_2:X_3$ (or $X_1:X_2:X_3$). Finally, the ligand $L$ binds to the signalling receptor $Z:X_1:X_2:X_3:X_4$ (or dummy receptor $X_1:X_2:X_3:X_4$), thus forming the signalling (or dummy) complex.}
  \label{fig:SRLK}
\end{figure}

The biochemical reaction network for a general SRLK model is given by
\begin{equation}
    \begin{array}{lllllccl}
    Z &+& X_1 &\rightleftharpoons& Z:X_1&&&K_0,\\
    X_2&+&Z:X_1 &\rightleftharpoons& Z:X_1:X_2&&&K_1,\\
    X_2&+&X_1 &\rightleftharpoons& X_1:X_2&&&K_1',\\
    \vdots&&\vdots&& \vdots&&&\vdots\\
    X_{i+1}&+&Z:X_1:\ldots :X_i &\rightleftharpoons& Z:X_1:\ldots :X_{i+1}&&&K_i,\\
    X_{i+1}&+&X_1:\ldots :X_i&\rightleftharpoons& X_1:\ldots :X_{i+1}&&& K_i',\\
    \vdots&&\vdots&& \vdots&&&\vdots\\
      X_{n}&+&Z:X_1:\ldots :X_{n-1} &\rightleftharpoons& Z:X_1:\ldots :X_{n}&&&K_{n-1},\\
    X_{n}&+&X_1:\ldots :X_{n-1}&\rightleftharpoons& X_1:\ldots :X_{n}&&& K_{n-1}',\\
    L&+&Z:X_1:\ldots :X_n&\rightleftharpoons& Z:X_1:\ldots :X_n:L &&&K_n,\\
    L&+&X_1:\ldots :X_n&\rightleftharpoons& X_1:\ldots :X_n:L &&&K_n',
\end{array}
\label{eq:SRLKcrn}
\end{equation}
where the $K_i$ (or $K_i'$) are the affinity constants related to the formation of the signalling (or dummy) complex. Figure~\ref{fig:SRLK}  illustrates the formation of the signalling and dummy complexes in an SRLK model with $n=4$ trans-membrane chains.
We assume the system  at steady state and that  ligand is in excess. 
In what follows We  refer to these
two assumptions as the \emph{experimental hypotheses}. 

We write $z$ (or $x_i$) for the steady state  concentration of unbound chain $Z$ (or $X_i$). We also use $L$ to denote the ligand concentration. Finally, $N_z$ (or $N_i$) denotes the total copy number per cell of the species $Z$ (or $X_i$). An SRLK model satisfying the experimental hypotheses is then described by the following polynomial system:
\begin{subequations}
    \begin{align}
    N_z&=z+K_0 zx_1+K_0K_1zx_1x_2+\ldots +K_0 K_1\ldots  K_{n-1}zx_1\ldots  x_n+K_0\ldots  K_nzx_1\ldots  x_nL\label{eq:polynomialSRLK_z}\\
    &=z+K_0z[x_1+\sum_{j=2}^n(\prod_{l=1}^{j-1}K_lx_lx_j)+L\prod_{j=1}^nK_jx_j],\nonumber\\
    N_1&=x_1 +K_1'x_1x_2+\ldots  +K_1'\ldots  K_{n-1}'x_1\ldots  x_n+K_1'\ldots  K_n' Lx_1\ldots  x_n\label{eq:polynomialSRLK_1}\\
    &\quad +K_0z(x_1+K_1x_1x_2+\ldots +K_1\ldots  K_{n-1}x_1\ldots  x_n+K_1\ldots  K_nx_1\ldots  x_nL),\nonumber\\
    \intertext{for $i=2,\ldots ,n-1$:}
    N_i&=x_i+K_1'\ldots  K_{i-1}'x_1\ldots x_i+\ldots +K_1'\ldots  K_{n-1}'x_1\ldots  x_n+K_1'\ldots  K_n'x_1\ldots  x_nL& \label{eq:polynomialSRLK_i}\\
    &\quad +K_0z(K_1\ldots  K_{i-1}x_1\ldots  x_i+\ldots +K_1\ldots  K_{n-1}x_1\ldots  x_n+K_1\ldots  K_n x_1\ldots  x_n L)\nonumber\\
    &=x_i+\sum_{j=i}^n(\prod_{l=1}^{j-1}K_l' x_l x_j+K_0z\prod_{l=1}^{j-1}K_lx_lx_j)+L\prod_{j=1}^nK_j'x_j+K_0zL\prod_{j=1}^n K_jx_j,\nonumber\\
    N_n&=x_n+K_0\ldots  K_{n-1}zx_1\ldots  x_n+K_1'\ldots  K_{n-1}'x_1\ldots  x_n+K_1'\ldots  K_n'L x_1\ldots  x_n+K_0\ldots  K_n z Lx_1\ldots  x_n
    \; .
    \label{eq:polynomialSRLK_n}
    \end{align}
    \label{eq:polynomialSRLK}
\end{subequations}
We note that 
many results in this section can be  further simplified  under the additional hypothesis of no allostery.
\begin{definition}
There is \emph{no allostery} in an SRLK model if $K_i=K_i'$ for all $i=1,\ldots ,n$.
\label{def:noallostery}
\end{definition}
Finally, we formally define
the signalling and dummy functions for this class of models.

\begin{definition}
For 
an SRLK model under the experimental hypotheses the signalling function, $\sigma(L)$, is the number of signalling complexes formed as a function of the ligand
concentration, $L$, and
can be written as follows
\[\sigma(L)=K_0zL\prod_{i=1}^nK_ix_i.\]
Similarly, the dummy function, $\delta(L)$, is the number of dummy complexes formed as a function of the 
ligand concentration, $L$,
and
can be written as follows
\[\delta(L)=L\prod_{i=1}^nK_i'x_i.\]
\label{def:SRLKsignalling}
\end{definition}
Note that the IL-7R model of Section~\ref{sec:il7r} is one example of  an SRLK model and the definition of signalling function given in Section~\ref{sec:amp-EC50} is equivalent.
We now introduce the 
 notion of a limiting component.

\begin{definition}
The species, $X_j$, which has the smallest total copy number of molecules $$0<N_j<N_i, \ \forall i\neq j,$$ is the limiting component of the system.
If there are multiple limiting components, $X_{j_1},\ldots ,X_{j_r}$, then $$0<N_{j_1}=\ldots =N_{j_r}<N_i, \ \forall i \notin \{j_1,\ldots ,j_r\}.$$
\label{def:limitingcomponent}
\end{definition}
If the signalling function attains its maximum for large values of the ligand concentration, then, since by definition $\sigma(0)=0$, the amplitude of such model is given by $$A\equiv\lim_{L\to+\infty}\sigma(L).$$
In this section we present some general results for
$\lim_{L\to+\infty}\delta(L)$
and
$\lim_{L\to+\infty}\sigma(L)$
applicable to
 SRLK models. The proofs of the lemmas and theorems can be found in Appendix~\ref{appendix:SRLK}.

\subsection{Asymptotic study of the steady states}
\label{sec:extra-properties}

While it is difficult to find closed-form expressions of the steady states for general receptor-ligand systems,
in what follows we show that considerable progress can be made for the specific case of SRLK models. In this section we describe the behaviour of the concentrations, $x_i$, in the limit $L\to+\infty$.
The proofs of our results can be found in Appendix~\ref{appendix:SRLK}. First, 
we recall the definition and a property of algebraic functions.

\begin{definition}
\label{def:defalgebraic}
A univariate function $y=f(x)$ is said to be algebraic if it satisfies the polynomial equation:
\begin{equation*} 
y^{m}+R_{m-1}(x)y^{m-1}+\cdots +R_{0}(x)=0, \tag{$\dagger$}\label{eq:alg_funct}
\end{equation*}
where the $R_i(x)$ are rational functions of $x$, {\em i.e.,} are of the form $\tfrac{p(x)}{q(x)}$, where $p$ and $q$ are polynomial functions and $q(x)\neq 0$ for all $x\in \mathbb{R}$.
\end{definition}
\begin{remark}
Note that the polynomial~\eqref{eq:alg_funct} has $m$ solutions. These solutions are called the branches of an algebraic function and one often specifies a particular branch.
\end{remark}
Since we are interested in the limit behaviour, the following lemma proves useful.
\begin{lemma}
\label{lem:algebraiclimit}
Any bounded, continuous solution of \eqref{eq:alg_funct} defined on $\mathbb{R}$ has a finite limit at $+\infty$ (and $-\infty$).
\end{lemma}
With this background in place, we can now proceed to study SRLK models in detail.
We start by showing that in steady state the signalling and  the dummy functions have a positive limit when $L$ tends to $+\infty$.
\begin{lemma}
\label{lem:sigdumlimit}
The signalling and the dummy functions of an SRLK model satisfying the experimental hypotheses admit a finite limit when $L\to +\infty$ and this limit is positive.
\end{lemma}
An equivalent result holds for the  steady state concentration of the kinase.
\begin{lemma}
In an SRLK model under the experimental hypotheses, the concentration  of the extrinsic intra-cellular kinase $Z$ admits a positive finite limit, $c_z>0$, when $L\to+\infty$.
\label{lem:JAKlim}
\end{lemma}
In the particular case of no allostery, we can write an explicit expression of the limit of $z$, $c_z$.
\begin{lemma}
Consider an SRLK model which satisfies the experimental hypotheses. If we assume no allostery, then the steady state value of the extrinsic intra-cellular kinase, $z$, is given by
\begin{align}
    z &=\frac{-1+K_1(N_z-N_1)+\sqrt{\Delta_z}}{2K_1},
    \label{eq:zvalue_noallost}
    \intertext{where}
    \Delta_z &=(1+K_1(N_1-N_z))^2+4K_1N_z.\nonumber
\end{align}
\label{lem:noallost_zindepL}
\end{lemma}
By Lemma~\ref{lem:noallost_zindepL} $z$ is independent of $L$ (thus, $c_z=z$) and only depends on $K_1$, $N_1$ and $N_z$. Note that this result is equivalent to the one obtained in Section~\ref{sec:il7r}
for the IL-7R model.
Finally, we study the behaviour of the concentration $x_i$ in the limit $L\to+\infty$. We first give bounds to the asymptotic dependency of $x_i$ on $L$.

\begin{lemma}
Let us consider an SRLK model which satisfies the experimental hypotheses.
Then no concentration $x_i$ behaves proportionally to $L^q$, $q>0$ or $\frac{1}{L^p}$, $p>1$ when $L\to+\infty$.
\label{lem:xidoesnotbehaveinLor1overLp}
\end{lemma}
We can now state the main theorem of this section.
\begin{theorem}
We consider an SRLK model which satisfies the experimental hypotheses. If there exists a unique limiting component $X_{i_0}$, then $$x_{i_0} \isEquivTo{L\to+\infty}\frac{c_{i_0}}{L},$$ and for all $i=1,  \ldots, n, i\neq i_0$, $$x_i\isEquivTo{L \to+\infty}c_i,$$ where $c_{i_0}$ and $c_i$ are positive constants.
\label{th:limitingcomponentbehaviour}
\end{theorem}

\begin{corollary}
 If an SRLK model, which satisfies the experimental hypotheses, has multiple limiting components, $X_{i_1},\ldots ,X_{i_r}$, $i_1<\ldots <i_r$, then 
 \[
 x_{i_1}\isEquivTo{L\to+\infty}
 \frac{c_{i_1}}{L^{p_1}},
 \ldots , x_{i_r}
 \isEquivTo{L\to+\infty}\frac{c_{i_r}}{L^{p_r}},\] where $c_{i_1},\ldots c_{i_r}$ are positive constants and $p_1=\ldots =p_r=\frac{1}{r}$. The
 concentrations of 
 the
 non-limiting components, $x_i$,  (for $i\notin\{i_1,\ldots ,i_r\}$) tend to positive constants, $c_i>0$. 
 \label{th:multiplelimitingcomponentbehaviour}
 \end{corollary}

\subsection{Asymptotic study of the signalling and dummy functions}
\label{sec:explicit-limit}

The previous section presented numerous small results which give insight into the steady state behaviour of SRLK receptor-ligand systems.
We are now in a position to combine these results to state and prove our main theorem, which gives closed-form formul\ae\ for the limits of the signalling and dummy functions.
\begin{theorem}
Consider an SRLK model which satisfies the experimental hypotheses. Let us write $X_{i_1},\ldots ,X_{i_r}$ as the limiting components and $N_{i_0}\equiv N_{i_1}=\ldots =N_{i_r}$ as their corresponding total number. 
The limit of the signalling function  is given by
\[\lim_{L\to+\infty}\sigma(L)=\frac{\prod_{i=1}^nK_iK_0c_z}{\prod_{i=1}^nK_i'+\prod_{i=1}^nK_iK_0c_z}N_{i_0},\]
and 
the limit of the dummy function  is
\[\lim_{L\to+\infty}\delta(L)=\frac{\prod_{i=1}^nK_i'}{\prod_{i=1}^nK_i'+\prod_{i=1}^nK_iK_0c_z}N_{i_0},\]
where $$c_z=\lim_{L\to+\infty}z.$$
\label{th:sigmadumlimexplicit}
\end{theorem}
Under the assumption of no allostery, these expressions can be
further simplified.
\begin{corollary}
Consider the system of Theorem~\ref{th:sigmadumlimexplicit} and assume there is no allostery. Denote the limiting components by $X_{i_1},\ldots ,X_{i_r}$ and $N_{i_0}\equiv N_{i_1}=\ldots =N_{i_r}$, their corresponding total number. The limit of the signalling function is
\[\lim_{L\to+\infty}\sigma(L)=\frac{K_0z}{1+K_0z}N_{i_0},\]
and 
the limit of the dummy function is
\[\lim_{L\to+\infty}\delta(L)=\frac{N_{i_0}}{1+K_0z},\]
with $z$ given by equation~\eqref{eq:zvalue_noallost} in Lemma~\ref{lem:noallost_zindepL}. 
\label{th:sigmadumlimexplicit_corr}
\end{corollary}
From the previous expressions
we observe that the limit of 
the signalling and dummy functions  are equal to the total copy number of  the limiting component, $N_{i_0}$, 
multiplied by a term which is bounded between $0$ and $1$. This term only depends on the affinity constant $K_0$ and the steady state concentration of the kinase.
In order to
relate the above limits back to biologically meaningful quantities, all there is left to show is that the explicit expression of the limit of $\sigma$ is in fact the amplitude of the system. Since $\sigma(0)=0$, let us first note that the amplitude is equal to the maximum of $\sigma$. Under the no allostery assumption, we can show mathematically that this maximum is the limit of $\sigma$ when $L\to+\infty$. To this end, the following lemma is needed.
\begin{lemma}
\label{lem:lim=amp_noallost}
Consider an SRLK model under the experimental hypotheses. If there is no allostery, then we have 
\[\sup \sigma(L)=\lim_{L\to+\infty}\sigma(L).\]
\end{lemma}
The supremum here is attained and is a maximum. Thus, the amplitude for a SRLK receptor-ligand system when there is no allostery is the limit of $\sigma$ described in Corollary~\ref{th:sigmadumlimexplicit_corr}. This result is 
the generalisation of the example discussed in Section~\ref{sec:il7r}.
We note that 
 the amplitude of the IL-7R model of Section~\ref{sec:il7r} can be recovered by setting $N_{i_0} =\min(N_x,N_y)$. We now have also rigorously shown that the limit of the signalling function is indeed the amplitude. The EC$_{50}$ can now be found as outlined in Section~\ref{sec:il7r}.

\subsection{SRLK models with additional sub-unit receptor chains}
\label{sec:SRLK-extensions}

As hinted in Section~\ref{sec:il7r+r}, the IL-7R model with the additional sub-unit receptor chain is part of a larger group of models which are an extension of the SRLK  family. Therefore, our previous results can be extended to this type of models. Again, we start by giving an abstract definition of the extended SRLK
family of models.

\begin{figure}[htp!]
    \centering
  \begin{subfigure}{\textwidth}
           \includegraphics[width=\textwidth]{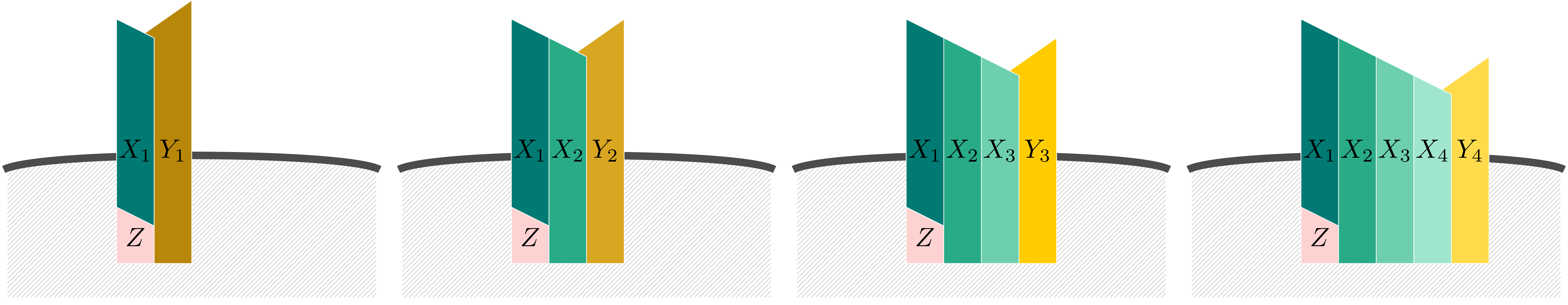}
           \caption{Decoy complexes with kinase.}
           \label{fig:srlk_decoyk}
  \end{subfigure}
  \begin{subfigure}{\textwidth}
           \includegraphics[width=\textwidth]{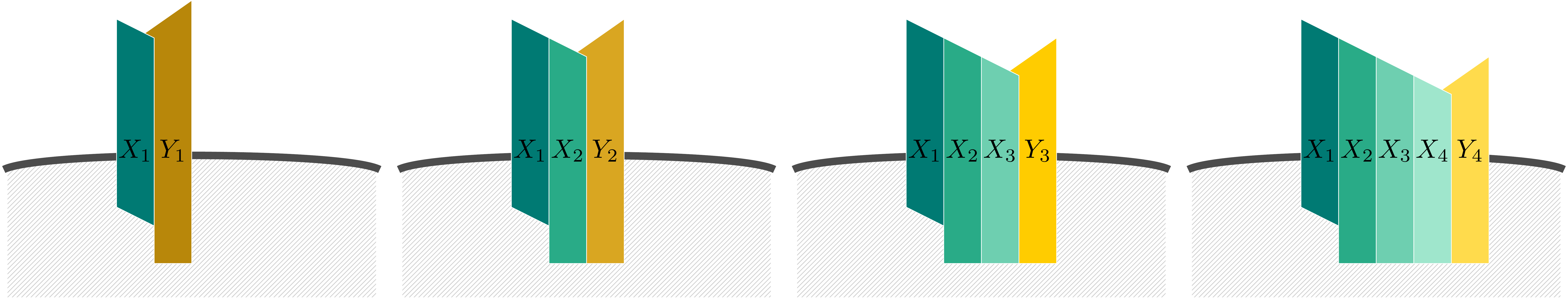}
           \caption{Decoy complexes without kinase.}
           \label{fig:srlk_decoynok}
  \end{subfigure}
  \begin{subfigure}{\textwidth}
           \includegraphics[width=\textwidth]{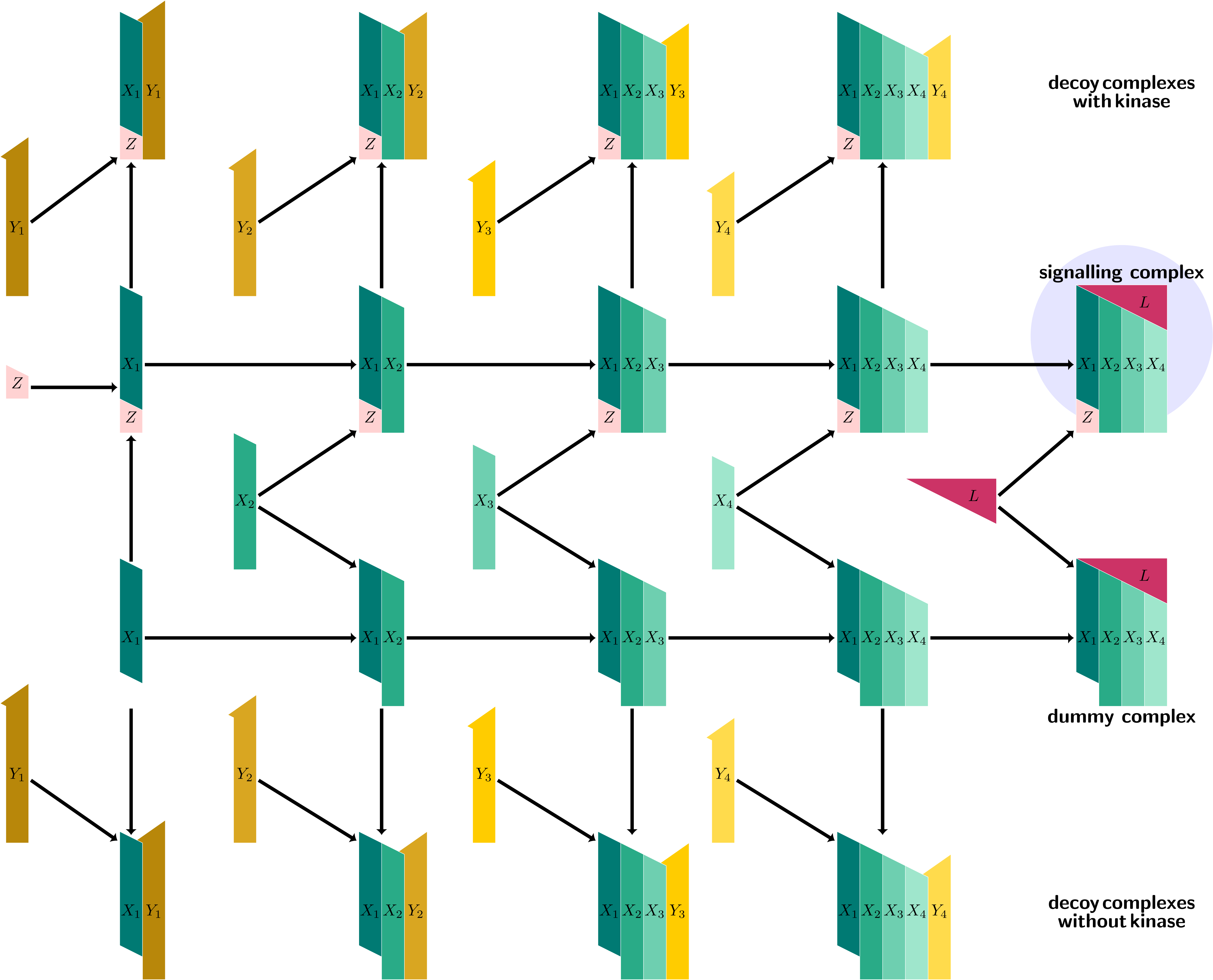}
           \caption{Extended SRLK model: sequential biochemical reaction scheme.}
           \label{fig:srlkext_scheme}
  \end{subfigure}
    \caption{Extended SRLK model with $n=4$ trans-membrane chains: (a) An additional sub-unit chain, $Y_i$, can bind to each intermediate signalling complex $Z:X_1:\ldots :X_i$, to form decoy complexes with kinase. (b) The sub-unit  chain $Y_i$ can also bind to the intermediate dummy complexes $X_1:X_2:\ldots :X_i$, forming decoy complexes without kinase. (c) Scheme of the sequential formation of the signalling and dummy complexes. At each step their formation can be interrupted by the binding of a sub-unit chain, $Y_i$, to the intermediate complex, forming a decoy complex. }
    \label{fig:SRLKext}
\end{figure}

\begin{definition}[Extended SRLK model]
An \emph{extended SRLK model} is an SRLK model where we assume that each intermediate complex, $Z:X_1:\ldots :X_i$ (or $X_1:\ldots :X_i$), for $i=1,\ldots ,n$  can bind to an extra chain, $Y_i$, with an affinity constant $K_{y_i}$ (or $K_{y_i}'$), to form a decoy complex $Z:X_1:\ldots X_i:Y_i$ (or $X_1:\ldots X_i:Y_i$). The addition of a sub-unit chain of the kind $Y_i$ prevents the binding of  ligand to the receptor, and thus,  does not allow the formation of signalling or dummy complexes.
\label{def:extensionSRLK}
\end{definition}
The chemical reaction network for an extended SRLK model is given by:
\begin{equation*}
    \begin{array}{lllllccl}
    Z &+& X_1 &\rightleftharpoons& Z:X_1&&&K_0\\
    Y_1&+&X_1&\rightleftharpoons&X_1:Y_1&&&K_{y_1}'\\
    Y_1&+&Z:X_1&\rightleftharpoons&Z:X_1:Y_1&&&K_{y_1}\\
    X_2&+&Z:X_1 &\rightleftharpoons& Z:X_1:X_2&&&K_1\\
    X_2&+&X_1 &\rightleftharpoons& X_1:X_2&&&K_1'\\
    Y_2&+&X_1:X_2&\rightleftharpoons&X_1:X_2:Y_2&&&K_{y_2}'\\
    Y_2&+&Z:X_1:X_2&\rightleftharpoons&Z:X_1:X_2:Y_2&&&K_{y_2}\\
    \vdots&&\vdots&& \vdots&&&\vdots\\
    X_{i+1}&+&Z:X_1:\ldots :X_i &\rightleftharpoons& Z:X_1:\ldots :X_{i+1}&&&K_i\\
    X_{i+1}&+&X_1:\ldots :X_i&\rightleftharpoons& X_1:\ldots :X_{i+1}&&& K_i'\\
    Y_i&+&X_1:\ldots :X_i&\rightleftharpoons&X_1:\ldots :X_i:Y_i&&&K_{y_i}'\\
    Y_i&+&Z:X_1:\ldots :X_i&\rightleftharpoons&Z:X_1:\ldots :X_i:Y_i&&&K_{y_i}\\
    \vdots&+&\vdots&& \vdots&&&\vdots\\
     Y_n&+&X_1:\ldots :X_n&\rightleftharpoons&X_1:\ldots :X_n:Y_n&&&K_{y_n}'\\
    Y_n&+&Z:X_1:\ldots :X_n&\rightleftharpoons&Z:X_1:\ldots :X_n:Y_n&&&K_{y_n}\\
    L&+&Z:X_1:\ldots .X_n&\rightleftharpoons& Z:X_1:\ldots :X_n:L &&&K_n\\
    L&+&X_1:\ldots .X_n&\rightleftharpoons& X_1:\ldots :X_n:L &&&K_n'
\end{array}
\end{equation*}
where $K_i$, $K_i'$, $K_{y_i}$ and $K_{y_i}'$ denote the affinity constants. Figure~\ref{fig:srlk_decoyk} and Figure~\ref{fig:srlk_decoynok} show the decoy complexes of an extended SRLK receptor-ligand system with $n=4$ trans-membrane chains. The signalling and dummy complexes are built sequentially similarly to the classic SRLK model (see Figure~\ref{fig:SRLK} and Figure~\ref{fig:srlkext_scheme}).

We note that while we assume  all the $X_i$ to be different species, we allow that $Y_i=Y_j$ or $Y_i=\emptyset$, as long as for $i=1,\ldots ,n$, $Y_i \notin \{X_1,\ldots ,X_n,Z,L\}$. 
We assume that the receptor-ligand system is in steady state and  the ligand is in excess. We further assume that the concentration of the species $Y_i$ (which we write $y_i$) are all bounded. We could consider
the case when the $Y_i$ are in excess, and thus, treat their concentration as a parameter of the model, or assume that the number
of $Y_i$ molecules is conserved. We refer to these assumptions as the \emph{extended experimental hypotheses}.

The signalling and dummy functions of classic and extended SRLK receptor-ligand
systems are equivalent (see Definition~\ref{def:SRLKsignalling}). The polynomial system describing an extended SRLK model under the extended experimental hypotheses is given by
\begin{subequations}
    \begin{align}
    N_z&=z+K_0z(x_1(1+K_{y_1}y_1)+\sum_{j=2}^n((1+K_{y_j}y_j)\prod_{l=1}^{j-1}K_lx_lx_j))+\sigma(L),\label{eq:polynomialextendedSRLK_z}\\
    N_1&=x_1(1+K_{y_1}'y_1) +\sum_{j=2}^n( (1+K_{y_j}'y_j)\prod_{l=1}^{j-1}K_l'x_lx_j)+\delta(L)\label{eq:polynomialextendedSRLK_1}\\
    &\quad +K_0z(x_1(1+K_{y_1}y_1)+\sum_{j=2}^n((1+K_{y_j}y_j)\prod_{l=1}^{j-1}K_lx_lx_j))+\sigma(L),\nonumber\\
    \vdots&\nonumber\\
    N_i&=x_i+\sum_{j=i}^n((1+K_{y_j}'y_j)\prod_{l=1}^{j-1}K_l'x_lx_j+K_0z(1+K_{y_j}y_j)\prod_{l=1}^{j-1}K_lx_lx_j)+\delta(L)+\sigma(L),\quad\text{for } i=2,\ldots ,n-1, \label{eq:polynomialextendedSRLK_i}\\
    \vdots&\nonumber\\
    N_n&=x_n+K_0z(1+K_{y_n}y_n)\prod_{j=1}^{n-1}K_jx_jx_n+(1+K_{y_n}'y_n)\prod_{j=1}^{n-1}K_j'x_jx_n+\delta(L)+\sigma(L).\label{eq:polynomialextendedSRLK_n}
    \end{align}
    \label{eq:polynomialextendedSRLK}
\end{subequations}
This system of polynomials is completed by the conservation equations of the species $Y_i$,
for $i=1,\ldots ,n$,
if we assume they are conserved. 

We can extend the notion of no allostery to the extended models.

\begin{definition}
An extended SRLK model is said to be under the assumption of \emph{no allostery} if for each $i=1,\ldots ,n$, $K_i=K_i'$ and $K_{y_i}=K_{y_i}'$. 
\end{definition}
With these expanded definitions, we can extend the results previously obtained for the SRLK receptor-ligand systems.
\begin{theorem}
\label{th:extensionSRLK}
The theorems and lemmas previously true for the SRLK models are true for the extended SRLK models under the same (extended) hypotheses.
\end{theorem}

\subsection{A few examples of (extended) SRLK models}
\label{sec:SRLKexamples}

In spite of some presumably strong modelling assumptions, the (extended) SRLK models can describe a broad range of 
cytokine-receptor  systems. The IL-7R models described in Section~\ref{sec:il7r} and  Section~\ref{sec:il7r+r}  are, respectively, an SRLK and an extended SRLK model. In this section, we provide examples of other interleukin-signalling systems which are part of the SRLK family.

\begin{example}[SRLK models: IL-2R and IL-15R]
The interleukin-2 (IL-2) receptor is composed of three trans-membrane 
sub-unit chains: the common gamma chain, $\gamma$, the IL-2R$\alpha$ chain and the IL-2R$\beta$ chain. Additionally, $\gamma$ binds to the intra-cellular extrinsic kinase JAK3. This IL-2 receptor-ligand system can be considered  an SRLK model with $\{Z,X_1,X_2,X_3,L\}=\{\text{JAK3},\gamma,\text{IL-2R}\beta,\text{IL-2R}\alpha, \text{IL-2}\}$. Similarly, the interleukin-15 (IL-15) receptor is composed of  three trans-membrane sub-unit chains, $\gamma$, IL-2R$\beta$ and IL-15R$\alpha$,
as well as the kinase JAK3. It can be considered an SRLK model with $\{Z,X_1,X_2,X_3,L\}=\{\text{JAK3},\gamma,\text{IL-2R}\beta,\text{IL-15R}\alpha,\text{IL-15}.\}$
\end{example}
A number of interleukin receptors share different molecular components. For instance, cytokine receptors of  the common gamma family (comprising the receptors for IL-2,4,7,9,15 and 21~\cite{rochman2009new}) share the common gamma chain, $\gamma$. In addition the IL-2 and IL-15 receptors share the sub-unit chain, IL-2R$\beta$.
The competition for these sub-unit chains can be mathematically described with an extended SRLK model, as follows.

\begin{figure}[htp!]
    \centering
     \begin{subfigure}[b]{0.4\textwidth}
        \includegraphics[width=\textwidth]{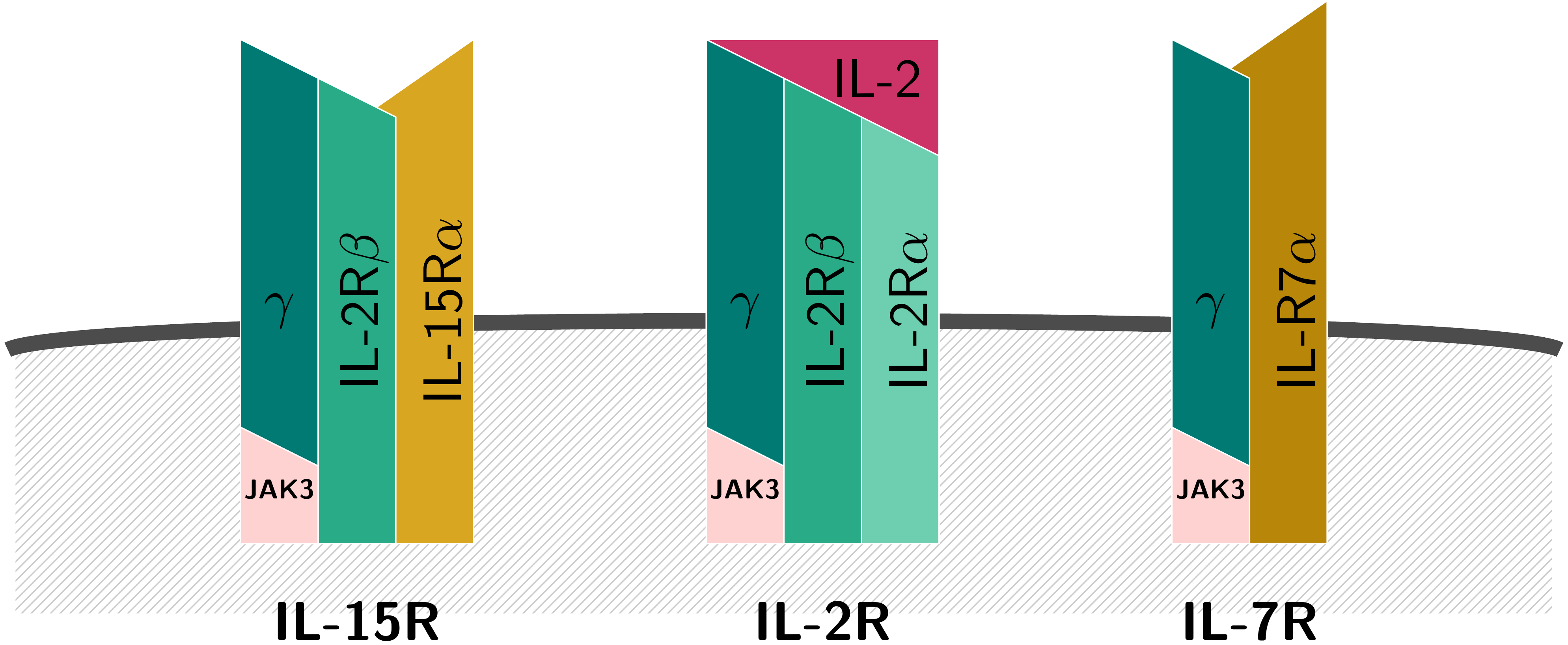}
         \caption{IL-2/IL-2R model with
         IL-7R and IL-15R competition.}
            \label{fig:gammafamily}
         \end{subfigure}
        \begin{subfigure}[b]{0.5\textwidth}
         \includegraphics[width=\textwidth]{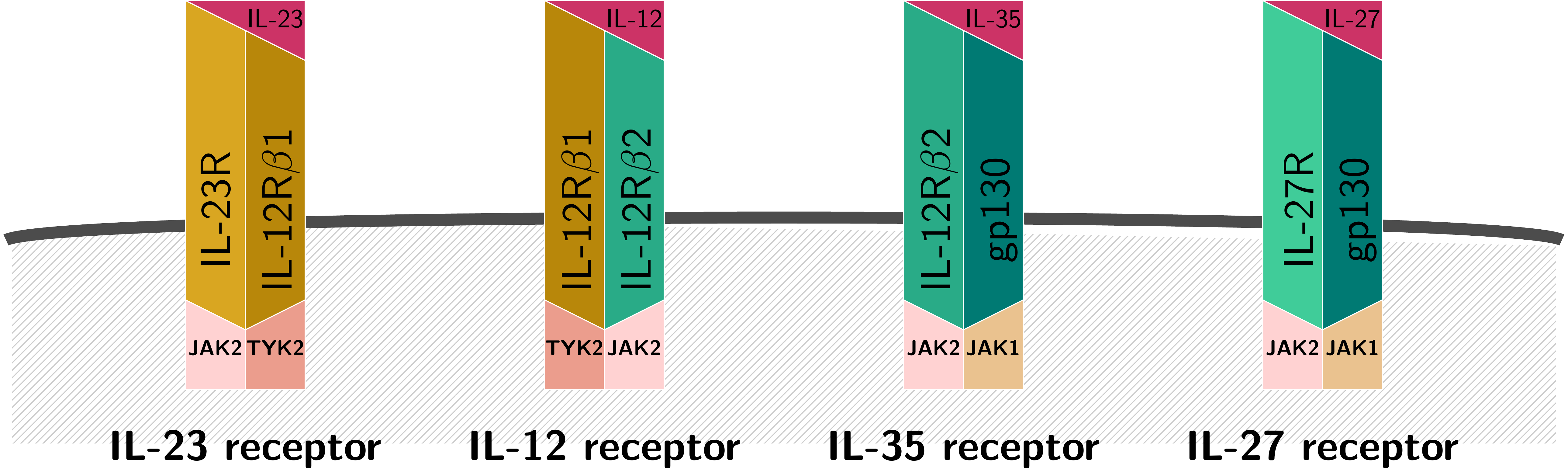}
         \caption{IL-12 receptor family.}
         \label{fig:il12ex}
         \end{subfigure}
    
      \caption{(a) Illustration of 
      example~\ref{ex:il2compet}: IL-2R, IL-7R and IL-15R competing for the common gamma chain and IL-2R$\beta$. The IL-2R is composed of three sub-unit chains: the gamma chain, IL-2R$\beta$ and IL-2R$\alpha$. IL-15R is composed of the gamma chain, IL-2R$\beta$ and the specific chain IL-15R$\alpha$. The IL-7R is composed of the gamma chain and IL-7R$\alpha$. All these receptors signal through the Janus kinase JAK3. (b) Illustration of example~\ref{ex:il12}:  (i) models the competition for IL-12R$\beta$1 between the IL-12 and the IL-23 receptors. We assume that IL-23R and IL-12R$\beta$2 are already bound to their associated extrinsic kinase JAK2;  (ii) models the competition for IL-12R$\beta$2 between the IL-12 and IL-35 receptors. We consider the complexes IL-12R$\beta$1:TYK2 and gp130:JAK1 already pre-formed;   (iii), models the competition for gp130 between the IL-35 and the IL-27 receptors. We consider the complexes IL-12R$\beta$2:JAK2 and IL-27R:JAK2 already pre-formed. }
    \label{fig:examples}
\end{figure}

\begin{example}[Extended SRLK model: IL-2/IL-2R model with formation of IL-7R and IL-15R]
\label{ex:il2compet}
Let us suppose we want to study the formation of IL-2/IL-2R complexes taking into account the competition for the $\gamma$ chain between IL-2R$\beta$ and IL-7R$\alpha$,  and the competition for the complex $\gamma:$IL-2R$\beta$ between
the sub-units IL-2R$\alpha$ and IL-15R$\alpha$. We can used an extended SRLK model with $$\{Z,X_1,X_2,X_3,L\}~=\{\text{JAK3},\gamma,\text{IL-2R}\beta,\text{IL-2}\alpha,\text{IL-2}\}$$ and $$\{Y_1,Y_2,Y_3\}=\{\text{IL-7R}\alpha,\text{IL-15R}\alpha,\emptyset\}.$$
This example is illustrated in Figure~\ref{fig:gammafamily}.
\end{example}

A further extended SRLK example is that of 
 the IL-12 family of receptors,
 which
 share multiple components~\cite{vignali201212},
 and
 each of which is
  composed of two trans-membrane
  sub-unit chains. The IL-12 receptor is composed of the 
  sub-unit chains IL-12R$\beta$1 and IL-12R$\beta$2. The IL-23 receptor signals via the IL-23R chain and
  the IL-12R$\beta$2 chain.
  The IL-27R (also known as WSX-1) and glycoprotein 130 (gp130) form the IL-27 receptor.
  Finally,  IL-12R$\beta$2 and gp130 form the IL-35 receptor.  The
  sub-unit chains gp130, IL-12R$\beta$1 and IL-12R$\beta$2 bind to a kinase from the JAK family (JAK1, TYK2 and JAK2 respectively). This competition can be described  with extended SRLK models as follows.

\begin{example}[Extended SRLK models: IL-12R family] 
\label{ex:il12} 
We provide three examples of extended SRLK systems which characterise the competition for receptor sub-units between  receptors of the IL-12 family (see Figure~\ref{fig:il12ex}).
\begin{enumerate}

    \item Suppose we want to study the IL-12-induced signalling process taking into account the competition for IL-12R$\beta$1. We can use an extended SRLK model with
\[\{Z,X_1,X_2,L,Y_1,Y_2\}=\{\text{TYK2},\text{IL-12R}\beta1,\text{IL-12R}\beta2^*,\text{IL-12},\text{IL-23R}^*,\emptyset\}.\]

\item 
To study IL-35-induced signalling taking into account the competition for IL-12R$\beta$2, we can use an extended SRLK model with \[\{Z,X_1,X_2,L,Y_1,Y_2\}=\{\text{JAK2},\text{IL-12R}\beta2,\text{gp130}^*,\text{IL-35},\text{IL-12R}\beta1^*,\emptyset\}.\]

\item An extended SRLK model with \[\{Z,X_1,X_2,L,Y_1,Y_2\}=\{\text{JAK1},\text{gp130},\text{IL-27R}^*,\text{IL-27},\text{IL-12R}\beta2^*, \emptyset\}\]
can describe the IL-27-induced signalling, when there is competition for the sub-unit chain gp130 with the IL-35 receptor.

\end{enumerate}
Above we have made use of the notation $X^*$  to denote the pre-formed complex composed of the receptor chain $X$ and its intra-cellular extrinsic kinase (TYK2 for IL-12R$\beta$1, JAK1 for gp130 and JAK2 for all the others).  
\end{example}


\section{Conclusion}
\label{sec:conclusion}

In the first part of this paper we propose a method to compute analytic expressions for two relevant pharmaco-dynamic metrics, the amplitude and  the EC$_{50}$ for receptor-ligand systems, based
on two (simple) IL-7 receptor models. Our method starts with the computation of a Gr\"obner basis for the polynomial system of the receptor-ligand system in steady state. As shown in our IL-7R models from Section~\ref{sec:il7r} and Section~\ref{sec:il7r+r}, 
the derivation of the amplitude
is easier when the maximum of the dose-response curve is attained at large ligand concentration (for instance when the dose-response curve is a sigmoid). In that case, the amplitude is the limit of the signalling function when the ligand concentration tends to infinity. When the model is simple enough, as is the case of the first IL-7R model, the polynomial system, simplified by the computation of the Gr\"obner basis, can be solved iteratively to obtain an analytic expression for the steady state. From these expressions, it is then relatively straightforward to compute the amplitude (\emph{i.e.,} the limit of the signalling function at large values of the ligand concentration) and the EC$_{50}$. For more complex models, such as our second IL-7R model, getting such steady state expressions can be more challenging. However, perturbation theory can be used to derive the expression for the amplitude. Computing another Gr\"obner basis can dramatically simplify the calculation of the EC$_{50}$, and in turn
display how it depends
on the parameters of the model. Analytic expressions for the amplitude and the EC$_{50}$ offer mechanistic insight for the receptor-ligand systems under consideration,  allow
one to quantify the parameter dependency of these two key variables, and can facilitate model validation and parameter exploration. Indeed, for both IL-7R models, we noticed that the affinity constant of the association of the gamma chain to the kinase JAK3, $K_1$, was the only constant involved in the expression for the amplitude. As a consequence, and if conducting parameter inference to fit the model to experimental data, $K_1$ would be the only parameter that could be inferred by comparison of the theoretical to the experimental amplitude. On the contrary, this constant was absent from both EC$_{50}$ expressions and thus, its value would be impossible to infer by only comparing the experimental 
to the theoretical EC$_{50}$. Our exact analysis also showed that both models have the same amplitude. Finally, the application of our method no longer requires the numerical computation
of the dose-response curve, finding its maximum to then obtain the amplitude and fit the curve to derive the EC$_{50}$. This reduces dramatically the computational cost and numerical errors. 
However, our method requires models simple enough to be able to compute a lex Gr\"obner basis, which is known to be computationally 
intensive~\cite{cox1997ideals,mityunin2007parallel}. Additionally, computing the amplitude when the maximum response is not the asymptotic behaviour of the dose-response curve can be tricky. For instance the computation of the maximum for bell-shaped dose-response curves (which has been done for simple models in Refs.~\cite{mack2008exact,douglass2013comprehensive}) may involve the computation of the derivative of the signalling function. This computation can be laborious even with the use of symbolic software. Finally, our method often requires additional mathematical tools or knowledge, such as perturbation theory in Section~\ref{sec:il7r+r}, which makes it rather a challenge to be used by those who are not mathematically trained. In
spite of the (sometimes, complicated) calculations that our method requires, we believe that  analytic expressions of the pharmacological metrics
characterising simple receptor-ligand systems may provide significant
advantages when
studying such biological systems. 

In the second part of this paper, we introduced a family of receptor-ligand systems, called SLRK, in which the signalling complex, composed of a kinase, a ligand and $n$ trans-membrane
sub-unit chains, is built sequentially. These models could also form dummy and decoy complexes, similarly to the IL-7R models which the SRLK family encompasses. By manipulating the polynomials describing the SRLK models, we are able
to derive an analytic expression of the amplitude under the no allostery assumption. We also show that the maximum of the dose response curve for both our IL-7R models was indeed the amplitude of the models. Despite relatively strong assumptions, we believe that the SRLK approach can be used to model a broad range of biochemical systems, such as receptor competition in interleukin signalling. The analytic expressions obtained for the amplitude could improve our understanding of biological mechanisms requiring a fine tuning of cytokine signalling such as cancer treatment~\cite{spolski2017gamma} or cytokine storm control~\cite{fajgenbaum2020cytokine,savarin2018fine}. 
We showed in Section~\ref{sec:SRLKexamples} how our SRLK models can account for the competition for the gamma chain between the IL-2 family of receptors and the competition for receptor components between the IL-12 family of receptors. However, many receptors signal through different
configurations. IL-35, for instance, can signal through homodimerisation of gp130 or IL-12R$\beta$2~\cite{collison2012composition}. It has been shown that IL-6, a cytokine implied in cytokine storms~\cite{chen2021soluble,fajgenbaum2020cytokine}, signals through an hexameric structure composed of two IL-6R$\alpha$ chains and two gp130 molecules~\cite{boulanger2003hexameric}. Furthermore, it seems that the ligand IL-6 first binds  to the IL-6R$\alpha$ chain before any association with gp130~\cite{boulanger2003hexameric}. Thus, one could imagine other general receptor models that may involve any of the following:
1) homo-oligomerisation (when two trans-membrane chains $X_i$ are identical), 2) other orders of signalling complex formation (non-sequential orders or for instance, if the ligand is not the final sub-unit to be bound), 
3) thermodynamic cycles (when there are several ways to form the signalling complex), 
4) multiple kinases (including kinases binding to other sub-unit chains, such as JAK1 which binds to IL-7R$\alpha$~\cite{park2019il7}), 
or 
5) a more detailed JAK-STAT pathway (most cytokine receptors activate multiple STAT
molecules, whose 
copy numbers tune the 
 immune response elicited~\cite{lin2019fine}). 

With this paper we hope to
have initiated, or renewed, an interest for the algebraic 
analysis of receptor-ligand systems.
Finally, we believe the results presented in this
paper are  
a first step to account 
for the variability of receptor expression levels when designing and studying receptor-ligand models (both from an experimental
and mathematical perspective)~\cite{farhat2021modeling,cotari2013cell,gonnord2018hierarchy,ring2012mechanistic}. 


\appendix

\section{Perturbation theory}
\label{appendix:perturbationtheory}

A well known difficulty with the lex G{\"o}bner basis method, and polynomial equations in general, is that there is 
usually no analytic solution when the degree of a univariate polynomial is greater than four. This result is known as the Abel-Ruffini theorem~\cite{zolkadek2000topological}. Therefore, in order to make progress, we need to resort to either numerical computations or analytic approximations. 
Since receptor-ligand systems are often
characterised by 
a sigmoidal dose-response curve,
at least to calculate the amplitude, the only quantity of interest is the limit of the signalling function at infinity.
In order to calculate this limit (where possible analytically, otherwise numerically) we make use of perturbation theory for polynomial equations.

Greatly inspired by the Dover book written by Simmonds and Mann~\cite{simmonds2013first}, this section reviews some notions of perturbation theory and justifies the steps of the method used to compute the analytic amplitude expression in Section~\ref{sec:il7r+r}. We start by defining an asymptotic expansion.
\begin{definition}[Asymptotic expansion]
We say that \[\sum_{n=0}^N \; c_n \; f_n(\epsilon),\] 
is an \textit{asymptotic expansion} of $f$ in $\epsilon$ if:
\begin{itemize}
    \item $\{f_n\}_{n=0, \ldots, N+1}$ is a \textit{gauge sequence}, {\em i.e.,} $f_n(\epsilon)=o(f_{n-1}(\epsilon))$ as $\epsilon\to 0$, for $n=0, \ldots, N+1$, and
    \item  $f(\epsilon)-\sum_{n=0}^N \; c_n \;  f_n(\epsilon)=\mathcal{O}(\epsilon^{N+1})$ as $\epsilon \to 0.$
\end{itemize}
\end{definition}

The core of perturbation theory is the notion of asymptotic expansion and the following fundamental theorem.
\begin{theorem}[Fundamental theorem of perturbation theory]
If an asymptotic expansion satisfies \[A_0+A_1\epsilon+A_2\epsilon^2+\ldots +A_n\epsilon^N+\mathcal{O}(\epsilon^{N+1})=0,\] for any sufficiently small $\epsilon$, and the coefficients $A_i$ are independent of $\epsilon$, then we have \[A_0=A_1=\ldots A_N=0.\]
\label{th:fundamentalTheoremPerturbation}
\end{theorem}

We are now ready to study the behaviour of the root of a univariate polynomial. Let $n \in \mathbb{N}^*$, where $\mathbb{N}^*$ is the set of natural numbers without zero. We consider a univariate polynomial, $P_{\epsilon}(x)$, of degree $n$, in the variable $x$, with coefficients which depend on the parameter $\epsilon$. We are interested in the behaviour of the roots of $P_{\epsilon} (x)$ when $\epsilon\to 0$. This polynomial can be re-written in the following form
\begin{equation}
    P_\epsilon(x)=(1+b_0\epsilon+c_0\epsilon^2+\ldots )+A_1\epsilon^{\alpha_1}(1+b_1\epsilon+c_1\epsilon^2+\ldots )x+\ldots +A_n\epsilon^{\alpha_n}(1+b_n\epsilon+c_n\epsilon^2+\ldots )x^n,
    \label{eq:originalPolynomialPerturbation}
\end{equation}
where 
for each $i$
$\alpha_i$ is a  rational number, $b_i$, $c_i$, $\ldots$ are real constants, $(1+b_i\epsilon+\ldots )$ 
is a regular asymptotic expansion of the general from \[a_0+a_1\epsilon+\ldots +a_N\epsilon^N+\mathcal{O}(\epsilon^{N+1}).\]
For such a polynomial, $P_{\epsilon} (x)$, we have the following result.
\begin{theorem}
\label{th:theoremroot}
Each root of a polynomial, such as equation \eqref{eq:originalPolynomialPerturbation} is of the form
\begin{equation}
    x(\epsilon)=\epsilon^p \omega(\epsilon), \ \ \ \omega(0)\neq 0,
\end{equation}
where $\omega$ is a continuous function of $\epsilon$  for $\epsilon$ sufficiently small and $p\in \mathbb{Q}$.
\end{theorem}

The proof of this theorem (see Ref.~\cite{simmonds2013first}) gives a method to study the asymptotic behaviour of the roots of 
polynomial \eqref{eq:originalPolynomialPerturbation}. 

\paragraph{Method.}
Let $P_\epsilon (x)$ be a polynomial that can be written as in equation \eqref{eq:originalPolynomialPerturbation}.
Let $p$ be a rational and $x$ a root of $P_{\epsilon}$. Let us replace $x$ by $\epsilon^p \omega(\epsilon)$ in $P_{\epsilon}$. We can re-write the polynomial as follows
\begin{equation}
    P_\epsilon(\epsilon^p \omega(\epsilon))=Q_\epsilon(\omega)+\epsilon(b_0+b_1A_1\epsilon^{\alpha_1+p}\omega(\epsilon)+\ldots +b_n A_n\epsilon^{\alpha_n+np}\omega(\epsilon))+\ldots,
    \label{eq:preparedPolynomialPerturbation}
\end{equation}
where \[Q_\epsilon(\omega)=1+A_1\omega(\epsilon)\epsilon^{\alpha_1+p}+\ldots A_n\omega(\epsilon)\epsilon^{\alpha_n+np}.\]
As $\epsilon \to 0$, the dominant term in $P_{\epsilon}$ is the term with the smallest exponent in $Q_\epsilon$, {\em i.e.,} the smallest element of
\begin{equation}
E=\{0,\alpha_1+p,\ldots ,\alpha_n+np\}  .  
\end{equation}
However, the set $E$ must have two identical values. Indeed, if $\alpha_k+kp$ is the smallest value of $E$, then we have
\[\epsilon^{-(\alpha_k+kp)}P_{\epsilon}(\epsilon^p\omega(\epsilon))\isEquivTo{\epsilon\to0}A_k \omega(0).\]
Since $\omega(0)\neq0$ and $P_{\epsilon}(\epsilon^p\omega(\epsilon))=0$ by hypothesis, we have $A_k=0$ which is a contradiction with the fact that $\alpha_k+kp\in E$. 
To select the proper value of $p$, we follow a graphical algorithm which indicates when two or more components of $E$ have equal minimal values:

\begin{enumerate}

    \item On a plane $(p,q)$, draw the lines $q=\alpha_j+j p$, $j=1,\ldots ,n$ and the line $q=0$.
    
    \item From the right, for $p$ sufficiently large, the smallest exponent is $0$. As $p$ decreases (one can imagine a fictive vertical line moving from right to left), there will be a first point where at least two lines intersect ($q=0$ and another one). Let us call this point $(p_1,0)$. One line will have the largest slope, $n_1$. 
    
    \item Let the fictive vertical line keep moving to the left and follow this line of slope $n_1$ until the next intersection $(p_2,e_2)$. Find the new intersected line with the largest slope $n_2$.
    
    \item Continue until there is no other intersection. The last intersection involves the line with the largest slope  of all the lines $n$. 
\end{enumerate}
We apply this method on an example and illustrate the algorithm in Section~\ref{sec:il7r+r}. This algorithm finds all the intersection points of the lines of equation $q=\alpha_j+jp$, $j=0, \ldots, n$ and $q=0$ that are on the lower envelop of these lines.
In this way, we have generated a set of pairs $\{(p_j,e_j)\}_{j=1, \ldots, m}$ corresponding to each intersection we encountered. Each of these intersection points corresponds to an asymptotic behaviour of one branch of the roots of our original polynomial $P_{\epsilon}$. Now let us define for each branch $j$, the scaled polynomial $T^{(j)}_{\epsilon}$, as follows.
\begin{equation}
    T^{(j)}_\epsilon(\omega)=\epsilon^{-e_j}P_\epsilon(\epsilon^{p_j}\omega).
\end{equation}
We can re-write $T^{(j)}_\epsilon$ as a sum of two polynomials
\[T_\epsilon^{(j)}(\omega)=T_0^{(j)}(\omega)+E_\epsilon^{(j)}(\omega),\]
where $E_0^{(j)}=0$ and $T_0^{(j)}$ do not  depend on $\epsilon$.
The non-zero roots of $T_\epsilon^{(j)}$ (approached by the roots of $T_0^{(j)}$ as $\epsilon\to0$) need to be regular but this is not necessarily the case. Indeed, $\alpha_j$ or $(p_j,e_j)$ may be non-integer rationals or $T_0^{(j)}$ may have repeated roots. To obtain regular expansions,, we introduce the new variable $\beta$ such that:
\begin{equation}
    \epsilon=\beta^{q_j},
\end{equation}
where $q_j$ is the least common denominator of the set $\{0, \alpha_1+p_j,\ldots \alpha_n+n p_j\}$. 
Finally, we define
\begin{equation}
    R_\beta^{(j)}(\omega)=T_j(\omega,\beta^{q_j})=\beta^{-q_je_j}P(\beta^{q_jp_j}\omega,\beta^{q_j})
\end{equation}
where $T_j(\omega,\epsilon)=T_\epsilon^{(j)}(\omega)$ and $P(\omega,\epsilon)=P_\epsilon(\omega)$.
The polynomial $R_\beta^{(j)}$ has the same roots as the polynomial $T_\epsilon ^{(j)}$ but its non-zero roots have a regular expansion in $\omega$ of the form
\[\omega(\beta)=a_0+a_1\beta+\ldots +a_N\beta^N+\mathcal{O}(\beta^{N+1}).\]
By substituting this expansion into $R_\beta^{(j)}$ and applying the fundamental theorem of perturbation theory 
(theorem~\ref{th:fundamentalTheoremPerturbation}), we find an expression for $a_0, a_1, \ldots$. 
We then come back to $x$ with the transformation $x=\beta^{q_jp_j}\omega(\beta)$ for each branch. In practice we explore each branch one by one and can eliminate those which are irrelevant (for instance when we have a negative root,
since in our case the roots of the polynomials are concentrations of species, or $\omega(0)=0$).

The above discussion can be summarised algorithmically as follows.

  \begin{center}
 \fbox{\parbox{\textwidth}{
\begin{enumerate}[leftmargin=20pt, rightmargin=20pt]

    \item Replace the variable $x$ by $\epsilon^p\omega(\epsilon)$ in $P_\epsilon(x)$, assuming $\omega(0)\ne 0$. 
    One obtains a polynomial of the form 
    \[P_{\epsilon}(\epsilon^p\omega(\epsilon))=Q_\epsilon(\omega)+\epsilon(\ldots )+\ldots .\]
    
    \item Write the set of exponents for $Q_\epsilon$:  $E=\{0,\alpha_1+p,\ldots ,\alpha_n+np\}$.
    
    \item Determine the pairs, $(p_j,e_j)$, of proper values and minimal exponents following the graphical algorithm described above. Each pair corresponds to an asymptotic branch to explore.

    \item For each branch $j$:
    \begin{enumerate}[label*=\arabic*.,leftmargin=20pt, rightmargin=20pt,noitemsep,nolistsep]
        \item Define $T^{(j)}_\epsilon(\omega)=\epsilon^{-e_j}P_\epsilon(\epsilon^{p_j}\omega)$.
        \item Introduce $\beta$ such that $\epsilon=\beta^q_j$, where $q_j=\text{lcd}(0,\alpha_1+p,\ldots ,\alpha_n+np)$, and define $R_\beta^{(j)}(\omega)=~T^{(j)}_{\beta^q_j}(\omega)$.
        \item In $R_\beta^{(j)}(\omega)=0$, substitute $\omega$ by a regular expansion $\omega(\beta)=a_0+a_1\beta+\ldots +a_N\beta^N+\mathcal{O}(\beta^{N+1}).$
        \item Apply the fundamental theorem of perturbation theory to obtain an analytic expression for $a_0, a_1,\ldots $. Usually at this step, we can discriminate whether this branch is relevant (see example~\ref{sec:il7r+r}).
        \item Find the asymptotic expansion of the root of the original polynomial, $P_\epsilon$, by $x=\beta^{q_jp_j}\omega(\beta).$
    \end{enumerate}
\end{enumerate}}}
\end{center}
In this paper we are mainly interested in the first non-zero coefficient of the regular expansion of $\omega$ 
since it drives the behaviour of the root of $P_\epsilon$ in the limit $\epsilon\to0$. 


\section{Computation of EC$_{50}$ for the IL-7R model} 
\label{appendix:il7r}

We make use 
of the expression for $\sigma(L)$, the signalling function described in \eqref{eq:il7rsig}, and equation \eqref{eq:il7r_ny-y}, 
to isolate the square root in equation \eqref{eq:il7r_ec50def}. 
\begin{equation}
    \sqrt{\Delta_2}=K_2(K_3L_{50}+1)(N_x+N_y)+1-\frac{K_2(K_3L_{50}+1)^2M}{K_3L},
    \label{eq:app_eq_EC50_1}
\end{equation}
with $M=\text{min}(N_x,N_y)$.
We square the equation to remove the root and simplify the expression to obtain
\begin{equation}
0=4K_2^2(K_3L_{50}+1)^2N_xN_y+K_2^2(K_3L_{50}+1)^4\frac{M^2}{K_3^2L_{50}^2}-2\frac{K_2^2(K_3L_{50}+1)^3M(N_x+N_y)}{K_3L_{50}}-2\frac{K_2(K_3L_{50}+1)^2M}{K_3L_{50}}.
\end{equation}
Since we are looking for a positive value of  $L_{50}$, we divide by $K_2(K_3L_{50}+1)^2$
and rewrite the previous expression as follows:
\begin{equation}
    0=4K_2K_3^2L_{50}^2N_xN_y+K_2(K_3L_{50}+1)^2M^2-2K_2(K_3L_{50}+1)M(N_x+N_y)K_3L_{50}-MK_3L_{50}.
\end{equation}
It leads to a polynomial of degree $2$ in $L_{50}$,
\begin{equation}
    0=M^2K_2+2K_3L_{50}M(-1+K_2(M-N_x-N_y))+K_3^2K_2L_{50}^2(M-2N_x)(M-2N_y).
\end{equation}
The discriminant of this polynomial is positive:
\begin{equation}
    \Delta=[1+K_2^2(N_y-N_x)^2+2K_2(N_x+N_y-M)]4K_3^2M^2,
\end{equation}
so that there are two potential solutions:
\begin{equation}
    \begin{aligned}
    L_{50}^+&=M\frac{1+K_2(N_x+N_y-M)+\sqrt{1+K_2^2(N_y-N_x)^2+2K_2(N_x+N_y-M)}}{K_2K_3(M-2N_x)(M-2N_y)},
    \\
    L_{50}^-&=M\frac{1+K_2(N_x+N_y-M)-\sqrt{1+K_2^2(N_y-N_x)^2+2K_2(N_x+N_y-M)}}{K_2K_3(M-2N_x)(M-2N_y)}.
    \end{aligned}
\end{equation}
Two solutions exist since by squaring
equation \eqref{eq:app_eq_EC50_1} we lose the positive steady state hypothesis. Substituting these expressions back into the steady state equations shows that only $L_{50}^+$ leads to a biologically relevant solution. The use of the algebraic method described at the end of Section~\ref{sec:il7r} is more elegant as it gives directly the correct EC50 expression.


\section{Macaulay2 code to compute Gr\"obner bases}
\label{appendix:M2code}

Every Gr\"obner basis of this paper has been computed making use
of  Macaulay2~\cite{M2}. We provide the code to compute the Gr\"oebner basis of the IL-7R model described in Section~\ref{sec:il7r}.

\begin{lstlisting}[language = python]
R = frac(QQ[Nx,Ny,Nz,L,K1,K2,K3])[x,y,z,MonomialOrder=>Lex]

I = ideal(- Nx + x + K2*x*y + K1*x*z + K2*K_1*x*y*z + K3*K2*L*x*y + K3*K2*K1*L*x*y*z,
- Ny + y + K2*x*y + K2*K_1*x*y*z + K3*K2*L*x*y + K3*K2*K1*L*x*y*z,
- Nz + z + K1*x*z + K2*K_1*x*y*z + K3*K2*K1*L*x*y*z)

g = gens gb I
\end{lstlisting}



\section{Analytic study of general sequential receptor-ligand systems}
\label{appendix:SRLK}

\subsection{Asymptotic study of the steady states}

\begin{replemma}{lem:algebraiclimit}
Any bounded, continuous solution of \eqref{eq:alg_funct} defined on $\mathbb{R}$ has a finite limit in $+\infty$ (and $-\infty$).
\end{replemma}

\begin{proof}
Multiply \eqref{eq:alg_funct} by the common denominator of the $R_i$ and let $x=\epsilon^{-1}$ to obtain $$\underbrace{\left[\prod_{i=0}^m \tilde{q}_i(\epsilon)\right]}_{\tilde{r}_m(\epsilon)} \tilde{y}^m + \tilde{r}_{m-1}(\epsilon)\tilde{y}^{m-1}+\dots+\tilde{r}_0(\epsilon)=0,$$ with $\tilde{y} = \tilde{f}(\epsilon)$. We have now recast the original problem into the form of equation \eqref{eq:originalPolynomialPerturbation}. By Theorem \ref{th:theoremroot} we know, that an expansion for the roots exists and we note that the points of $f(x)$ as $x\to\infty$ correspond to the points of $\tilde{f}(\epsilon)$ as $\epsilon=0$.
Note that, since all real $f(x)$ are bounded, so are the real $\tilde{f}(\epsilon)$. Therefore all real $\tilde{f}(0)$ are finite and equal to the limits $\lim_{x\to\infty}f(x)$. A unique limit is chosen by specifying a branch of $f(x)$. The proof for $x\to -\infty$ follows mutatis mutandis.
\end{proof}

\begin{replemma}{lem:sigdumlimit}
The signalling function $\sigma$ and the dummy function $\delta$ of an SRLK model satisfying the experimental hypotheses admit a finite limit when $L\to +\infty$ and this limit is positive.
\end{replemma}

\begin{proof}
The function $\sigma$ (or $\delta$) are algebraic functions bounded on $\mathbb{R}$ between $0$ and $\text{min}(N_z, N_1,\ldots, N_n)$ (or $\text{min}(N_1, \ldots, N_n)$) so they admit a finite limit when $L\to+\infty$. Let us denote this limit by $c_{\sigma}$ (or $c_{\delta}$). We know that $c_{\sigma}$ and $c_{\delta}$ are non-negative because $\sigma$ and $\delta$ are products of non-negative functions.

Consider $c_{\delta}=0$. Then since $\sigma(L)=K_0z\prod_{i=1}^n\frac{K_i}{K_i'}\delta(L)$,  we have $c_{\sigma}=0$ (we note that $z$ being also an algebraic function, $z$ also admits a finite limit when $+\infty$). Since $\delta$ converges to $0$, we need 
\begin{equation}
\prod_{i=1}^nx_i\isEquivTo{L\to+\infty}\frac{C_n}{L^p},
\label{eq:limitprodxi}
\end{equation}with $C_n$ a positive constant and $p>1$. We recall and rewrite polynomial \eqref{eq:polynomialSRLK_n}:
\begin{equation}
N_n=x_n+K_0z\prod_{i=1}^{n-1}K_i\prod_{i=1}^nx_i+\prod_{i=1}^{n-1}K_i'\prod_{i=1}^nx_i+\delta(L)+\sigma(L).
\label{eq:polySRLK_n_proof}
\end{equation}
Assuming \eqref{eq:limitprodxi} when $L\to+\infty$ in \eqref{eq:polySRLK_n_proof}, we obtain:
\[\lim_{L\to+\infty}x_n=N_n,\]
and so we must have $$\prod_{i=1}^{n-1}x_i\isEquivTo{L\to+\infty}\frac{C_{n-1}}{L^p},$$ with $p>1$ and $C_{n-1}$ a positive constant. Passing to the limit in polynomial \eqref{eq:polynomialSRLK_i} for $i=n-1$, we obtain $$\lim_{L\to+\infty}x_{n-1}=N_{n-1}.$$
We repeat the process for every conservation equation \eqref{eq:polynomialSRLK_i} of the species $X_i$ and we obtain
\[\forall i=1,\ldots, n, \ \lim_{L\to+\infty}x_i=N_i,\] which is a contradiction with equation \eqref{eq:limitprodxi}.  So $c_\delta>0$. \\

Now, consider $c_{\sigma}=0$. Then since $\sigma(L)=K_0z\prod_{i=1}^n\frac{K_i}{K_i'}\delta(L)$, $z$ has to tend to $0$. However, when passing to the limit $L\to+\infty$ in equation \eqref{eq:polynomialSRLK_z}, we obtain $$N_z=\lim_{L\to+\infty}(z+K_0 z x_1+\ldots +\sigma(L))=0,$$
which is a contradiction.

Conclusion: $c_{\sigma}>0$ and $c_{\delta}>0$.
\end{proof}

\begin{replemma}{lem:JAKlim}
In an SRLK model under the experimental hypotheses, the concentration $z$ of the intra-cellular extrinsic kinase $Z$ admits a positive finite limit $c_z>0$ when $L\to+\infty$.
\end{replemma}

\begin{proof}
The concentration of kinase $z$ being an algebraic function bounded on $\mathbb{R}$ between $0$ and $N_z$, it admits a finite limit $c_z$ when $L\to+\infty$. We know that $c_z \geq 0$ because $z$ is a concentration. We now prove that $c_z>0$.
Since $\delta$ converges to a positive constant when $L\to+\infty$, we must have $$\prod_{i=1}^nx_i\isEquivTo{L\to+\infty}\frac{c_d}{L},$$ where $c_d$ is a positive constant. Since $\sigma$ also admits a finite limit when $L\to+\infty$, it means that $$z\prod_{i=1}^nx_i\isEquivTo{L\to+\infty}\frac{c_s}{L},$$ where $c_s$ is a positive constant. So $z$ has to satisfy \[z\isEquivTo{L\to+\infty}c_z,\]
where $c_z=\frac{c_s}{c_d}$ is a positive constant.
\end{proof}

\begin{replemma}{lem:noallost_zindepL}
Consider an SRLK model which satisfies the experimental hypotheses. If we assume no allostery, then the steady state value of $z$ is given by
\begin{align}
    z &=\frac{-1+K_1(N_z-N_1)+\sqrt{\Delta_z}}{2K_1},
    \label{eq:zvalue_noallost_appendix}
    \intertext{where}
    \Delta_z &=(1+K_1(N_1-N_z))^2+4K_1N_z.\nonumber
\end{align}
\end{replemma}

\begin{proof}
We assumed no allostery so $K_i=K_i'$ for all $i=1,\ldots ,n$.
Equation \eqref{eq:polynomialSRLK_z} gives:
\[N_z-z=K_0z
\left(
x_1+\sum_{j=2}^n
\left( \prod_{l=1}^{j-1}K_lx_lx_j \right) +L\prod_{j=1}^nK_jx_j
\right) \]
By substituting this equality in equation \eqref{eq:polynomialSRLK_1}, we obtain:
\[N_{1}=N_z-z+\frac{N_z-z}{K_0z},\]
so $z$ is a positive root of the polynomial 
\[-N_z+z(1+K_1(N_1-N_z))+K_1z^2,\]
with $L$-independent coefficients.
The two possibilities are:
\begin{align*}
    z_1&=\frac{-1+K_1(N_z-N_1)+\sqrt{4K_1N_z+(1+K_1(N_1-N_z))^2}}{2K_1},
    \\
    z_2&=\frac{-1+K_1(N_z-N_1)-\sqrt{4K_1N_z+(1+K_1(N_1-N_z))^2}}{2K_1}.
\end{align*}
The expression $z_1$ is always positive while $z_2$ is always negative. Hence $z_1$ is the steady state kinase concentration, $z$.
\end{proof}

\begin{replemma}{lem:xidoesnotbehaveinLor1overLp}
Let us consider an SRLK model which satisfies the experimental hypotheses.
Then no concentration $x_i$ behaves proportionally to $L^q$, $q>0$ or $\frac{1}{L^p}$, $p>1$ when $L\to+\infty$.
\end{replemma}
\begin{proof}
Lemma \ref{lem:JAKlim} affirms that $z$ tends to a positive constant when $L\to+\infty$. In order for $\sigma$ or $\delta$ to converge to a positive constant as stated in lemma \ref{lem:sigdumlimit}, we need \begin{equation}\prod_{i=1}^nx_i\isEquivTo{L\to+\infty}\frac{c}{L},
\label{eq:prodxi} \end{equation} where $c$ is a positive constant. Since the concentrations $x_1,\ldots ,x_n$ are bounded functions (between 0 and their respective $N_{i}$), it is impossible to have for any $i=1\ldots n$, $x_i\isEquivTo{L\to+\infty}c_iL^q$ with $c_i$ constant and $q>0$. From equation \eqref{eq:prodxi} it follows that it is impossible to have any $x_i\isEquivTo{L\to+\infty}\frac{c_i}{L^p}$ for $p>1$.
\end{proof}

\begin{reptheorem}{th:limitingcomponentbehaviour}
We consider an SRLK model which satisfies the experimental hypotheses. If there exists a unique limiting component $X_{i_0}$ then $$x_{i_0} \isEquivTo{L\to+\infty}\frac{c_{i_0}}{L},$$ and for all $i=1, \ldots, n, i\neq i_0$, $$x_i\isEquivTo{L \to+\infty}c_i,$$ where $c_{i_0}$ and  $c_i$ are positive constants.
\end{reptheorem}

\begin{proof}
Since the concentrations $x_i$ are algebraic functions (with coefficients in $\RR$) bounded on $\mathbb{R}$, they admit a non-negative limit when $L\to+\infty$. 

We know that we need
\begin{equation}
\prod_{i=1}^nx_i\isEquivTo{L\to +\infty}\frac{c}{L},
\label{eq:proof_prodxiconverges}
\end{equation}
with $c$ a positive constant, so that $\sigma$ and $\delta$ converge when $L\to+\infty$. Lemma \ref{lem:JAKlim} shows that $z$ tends to a positive constant when $L\to+\infty$. Thus, it follows from equation \eqref{eq:proof_prodxiconverges} and Lemma \ref{lem:xidoesnotbehaveinLor1overLp} that at least one of the $x_i$ must tend to $0$. We will prove that the only concentration that can tend to $0$ is $x_{i_0}$ and so $x_{i_0}\isEquivTo{L\to+\infty}\frac{c_{i_0}}{L}$,  with $c_{i_0}$
a constant.

1) There exists at least one chain $X_j$ whose concentration tends to $0$. The conservation equation of $X_j$ described in equation \eqref{eq:polynomialSRLK_i} is:
\[N_j=x_j+\prod_{i=1}^jK_i'x_i+K_0z\prod_{i=1}^jK_ix_i+\prod_{i=1}^{j+1}K_i'x_i+K_0z\prod_{i=1}^{j+1}K_ix_i+\ldots +\delta(L)+\sigma(L).\]
When $L\to +\infty$, we obtain 
\[N_j=\lim_{L\to+\infty}\delta(L)+\lim_{L\to+\infty}\sigma(L).\]
We cannot form more dummy or signalling complexes than 
the number of molecules available.
Since $X_{i_0}$ is the limiting component, we have 
$$\delta(L)+\sigma(L)\leq N_{i_0}, \ \ \forall L.$$
This yields in the limit  $L\to+\infty$, $N_j\leq N_{i_0}$. By hypothesis this implies means that $j=i_0$ and so $X_j$ is our limiting component $X_{i_0}$.

2) Reciprocally, if $x_{i_0}$ tends to a positive constant when $L\to+\infty$, then there exists at least one $x_j$, $j\neq i_0$ such that $x_j\to 0$ when $L\to+\infty$. The limit when $L\to+\infty$ of equation \eqref{eq:polynomialSRLK_i} when $i=j$ gives
\[\lim_{L\to+\infty}[\delta(L)+\sigma(L)]=N_j.\]
However, since we also have $\delta+\sigma\leq N_{i_0}$, we obtain when taking the limit, $N_j \leq N_{i_0}$, which is a contradiction with the fact that $X_{i_0}$ is the only limiting component.

Conclusion:  $X_{i_0}$ is limiting if and only if its concentration tends to $0$, and we have
$$x_{i_0}\isEquivTo{L\to+\infty}\frac{c_{i_0}}{L},$$ and for $i\neq i_0$, $$x_{i}\isEquivTo{L\to+\infty}c_i,$$
where $c_{i_0}$ and $c_{i}$ are positive constants. 
 \end{proof}
 
 \begin{repcorollary}{th:multiplelimitingcomponentbehaviour}
  If an SRLK model, which satisfies the experimental hypotheses, has multiple limiting components $X_{i_1},\ldots ,X_{i_r}$, $i_1<\ldots <i_r$ then 
 \[x_{i_1}\isEquivTo{L\to+\infty}\frac{c_{i_1}}{L^{p_1}},\ldots ,x_{i_r}\isEquivTo{L\to+\infty}\frac{c_{i_r}}{L^{p_r}},\] where $c_{i_1},\ldots c_{i_r}$ are positive constants and $p_1=\ldots =p_r=\frac{1}{r}$. The non-limiting components 
 $x_i$ tend to positive constants $c_i$ with $i\notin\{i_1,\ldots ,i_r\}$. 
 \end{repcorollary}
 
  \begin{proof}
 If $X_{i_1}$ and $X_{i_2}$ are limiting components, they are the only ones whose concentrations, $x_{i_1}$ and $x_{i_2}$, tend to $0$ when $L\to+\infty$. From equation \eqref{eq:proof_prodxiconverges}
 we can write
 \begin{equation*}
     \begin{aligned}
     x_{i_1}&\isEquivTo{L_\to+\infty}\frac{c_{i_1}}{L^{p_1}},
     \\
      x_{i_2}&\isEquivTo{L_\to+\infty}\frac{c_{i_2}}{L^{p_2}},
     \end{aligned}
 \end{equation*}
with $c_{i_1}$ and $c_{i_2}$ constants and $p_1,p_2>0$, such that $p_1+p_2=1$. 

From system \eqref{eq:polynomialSRLK}, we have:
\begin{equation*}
    \begin{aligned}
    N_{i_1}&=x_{i_1}+\sum_{j=i_1}^n(\prod_{l=1}^{j-1}K_l'x_l x_j+K_0z\prod_{l=1}^{j-1}K_lx_lx_j)+L\prod_{j=1}^nK_j'x_j+K_0zL\prod_{j=1}^n K_jx_j,
    \\
    N_{i_2}&=x_{i_2}+\sum_{j=i_2}^n(\prod_{l=1}^{j-1}K_l'x_l x_j+K_0z\prod_{l=1}^{j-1}K_lx_lx_j)+L\prod_{j=1}^nK_j'x_j+K_0zL\prod_{j=1}^n K_jx_j.
    \end{aligned}
\end{equation*}
Since $X_{i_1}$ and $X_{i_2}$ are limiting components, we have $N_{i_1}=N_{i_2}$ and, if $i_1<i_2$, we obtain
 \begin{equation}
     \begin{aligned}
     N_{i_1}=N_{i_2}&\leftrightarrow&x_{i_1}(1+\sum_{j=i_1}^{i_2-1}(\prod_{l=1}^{j-1} K_l' \prod_{l=1, l\neq i_1}^{j} x_l +K_0z\prod_{l=1}^{j-1} K_l \prod_{l=1, l\neq i_1}^{j}x_l ))=x_{i_2}.
     \end{aligned}
     \label{eq:ni1=ni2}
 \end{equation}
Since all the $x_i$, with $i\neq i_1$, $i\neq i_2$, tend to a positive constant when $L\to+\infty$, we have  \[1+\sum_{j=i_1}^{i_2-1}(\prod_{l=1}^{j-1} K_l' \prod_{l=1, l\neq i_1}^{j} x_l +K_0z\prod_{l=1}^{j-1} K_l \prod_{l=1, l\neq i_1}^{j}x_l) \isEquivTo{L\to+\infty}C,\] 
where $C$ is a positive constant. Thus, we obtain the behaviour of the left side of equation \eqref{eq:ni1=ni2}
 \begin{equation*}
     x_{i_1}(1+\sum_{j=i_1}^{i_2-1}(\prod_{l=1}^{j-1} K_l' \prod_{l=1, l\neq i_1}^{j} x_l +K_0z\prod_{l=1}^{j-1} K_l \prod_{l=1, l\neq i_1}^{j}x_l ))\isEquivTo{L_\to+\infty}\frac{Cc_{i_1}}{L^{p_1}}.
 \end{equation*}
Since the right side is given by
 \begin{equation*}
        x_{i_2}\isEquivTo{L\to+\infty}\frac{c_{i_2}}{L^{p_2}},
 \end{equation*}
it results in $p_1=p_2=\frac{1}{2}$. 
 
 If there are $r$ limiting components $x_{i_1},\ldots ,x_{i_r}$, $i_1<\ldots <i_r$, then we have
  \begin{equation*}
     \begin{aligned}
     x_{i_1}&\isEquivTo{L_\to+\infty}&\frac{c_{i_1}}{L^{p_1}},
     \\
     \vdots&&\vdots\\
      x_{i_r}&\isEquivTo{L_\to+\infty}&\frac{c_{i_r}}{L^{p_r}},
     \end{aligned}
 \end{equation*}
with $c_{i_1},\ldots ,c_{i_r}$  positive constants, and $p_1,\ldots ,p_r>0$ such that $p_1+ \ldots+p_r=1$.
 We proceed the same way as for the case of two limiting components and we obtain $p_1=\ldots =p_r= \frac{1}{r}$. 
 \end{proof}
 
 \subsection{Asymptotic study of the signalling and dummy functions}
 
 \begin{reptheorem}{th:sigmadumlimexplicit}
 Consider an SRLK model which satisfies the experimental hypotheses. Write $X_{i_1},\ldots ,X_{i_r}$ as the limiting components and $N_{i_0}\equiv N_{i_1}=\ldots =N_{i_r}$ as their corresponding total number per cell. 
The limit of the signalling function $\sigma$ when $L$ tends to
$+\infty$ is
\[\lim_{L\to+\infty}\sigma(L)=\frac{\prod_{i=1}^nK_iK_0c_z}{\prod_{i=1}^nK_i'+\prod_{i=1}^nK_iK_0c_z}N_{i_0}.\]
The limit of the dummy function $\delta$  when $L$ tends to
$+\infty$ is
\[\lim_{L\to+\infty}\delta(L)=\frac{\prod_{i=1}^nK_i'}{\prod_{i=1}^nK_i'+\prod_{i=1}^nK_iK_0c_z}N_{i_0},\]
where $$c_z=\lim_{L\to+\infty}z.$$
 \end{reptheorem}
 
 \begin{proof}
By definition of $\sigma$ and $\delta$ we have:
 $$\delta(L)+\sigma(L)=L\prod_{i=1}^nx_i
 \left( \prod_{i=1}^nK_i'+K_0z\prod_{i=1}^nK_i \right),$$
which implies that
$$\lim_{L\to+\infty}[\delta(L)+\sigma(L)]=\lim_{L\to+\infty}
\left( L\prod_{i=1}^nx_i \left( \prod_{i=1}^nK_i'+K_0z\prod_{i=1}^nK_i \right)\right).$$
Using the limit properties and since everything converges, we obtain:
$$\lim_{L\to+\infty}[\delta(L)+\sigma(L])=\lim_{L\to+\infty}
\left( L\prod_{i=1}^nx_i \right)\left( \prod_{i=1}^nK_i'+K_0\lim_{L\to\infty}(z)\prod_{i=1}^nK_i \right).$$ 
However, theorem \ref{th:limitingcomponentbehaviour} states that $x_{i_0}$ tends to $0$ when $L\to+\infty$. Thus, equation \eqref{eq:polynomialSRLK_i} when $i=i_0$ at $L\to+\infty$ gives $$N_{i_0}=\lim_{L\to+\infty}[\delta(L)+\sigma(L)].$$
Consequently since $z\to c_z>0$ from lemma~\ref{lem:JAKlim}, we obtain:
 $$\lim_{L\to+\infty}(L\prod_{i=1}^nx_i)=\frac{N_{i_0}}{\prod_{i=1}^nK_i'+K_0c_z\prod_{i=1}^nK_i}.$$
 We substitute this limit into the expression of $\sigma$ and $\delta$ and obtain the desired expressions. 
\end{proof} 

\begin{repcorollary}{th:sigmadumlimexplicit_corr}
 Consider the system of Theorem~\ref{th:sigmadumlimexplicit} and assume there is no allostery. Denote the limiting components by $X_{i_1},\ldots ,X_{i_r}$ and $N_{i_0}\equiv N_{i_1}=\ldots =N_{i_r}$, their corresponding total number per chain. The limit of the signalling function $\sigma$ at $+\infty$ is
\[\lim_{L\to+\infty}\sigma(L)=\frac{K_0z}{1+K_0z}N_{i_0}.\]
The limit of the dummy function $\delta$ at $+\infty$ is
\[\lim_{L\to+\infty}\delta(L)=\frac{N_{i_0}}{1+K_0z},\]
and $z$ given by equation \eqref{eq:zvalue_noallost} in Lemma~\ref{lem:noallost_zindepL}. 
\end{repcorollary}

\begin{proof}
Since there is no allostery, we have $K_i=K_i'$ for all $i$. Lemma~\ref{lem:noallost_zindepL} states that  $z$ is independent of $L$, thus $c_z=z$. Applying these statements in the expressions of the previous theorem, we obtain the expressions of this corollary. 
\end{proof}

\begin{replemma}{lem:lim=amp_noallost}
Consider an SRLK model under the experimental hypotheses. If there is no allostery, then we have 
\[\sup \sigma=\lim_{L\to+\infty}\sigma(L).\]
\end{replemma}
\begin{proof}
Since $X_{i_0}$ is the limiting component, we know from Theorem~\ref{th:limitingcomponentbehaviour} that its concentration tends to $0$ when $L\to+\infty$. We have
\begin{equation}
    \delta+\sigma\leq N_{i_0}=\lim_{L\to+\infty}(\delta+\sigma).
    \label{eq:sig+del}
\end{equation}
In the no allostery case, $z$ is independent of $L$ and we have $\sigma=K_0z\delta$. Thus, equation \eqref{eq:sig+del} gives
\[(1+K_0z)\delta \leq (1+K_0z)\lim_{L\to+\infty}\delta.\]
Hence, we can conclude
\[\lim_{L\to+\infty}\delta=\sup\delta,\]
and \[\lim_{L\to+\infty}\sigma=\sup\sigma.\]


\end{proof}

\subsection{SRLK models with additional receptor sub-units}

\begin{reptheorem}{th:extensionSRLK}
The theorems and lemmas previously true for the SRLK models are true for the extended SRLK models under the same (extended) hypotheses.
\end{reptheorem}

\begin{proof}
The concentrations $y_i$ are bounded ($0\leq y_i\leq N_{y_i}$) algebraic function on $\mathbb{R}$, and therefore admit a limit when $L\to+\infty$. As the expressions of $\sigma$ and $\delta$ are not modified, the addition of the $Y_i$ variables to a SRLK model, assuming the extended experimental hypotheses, does not change the proofs of the previous lemmas and theorems. 
\end{proof}

\enlargethispage{20pt}





\paragraph{Funding.} This project has received funding from the European Union's Horizon 2020 research and innovation programme under the Marie Sk\l{}odowska-Curie grant agreement number 764698 (LS and CMP).

\paragraph{Acknowledgements:} We would like to thank Gr\'egoire Altan-Bonnet for inspiring this work and for encouraging us to explore with analytical methods how the dose-response curve depends on receptor expression levels. 
We thank Elisenda Feli\'u
and Alicia Dickenstein
for carefully reading
and providing detailed feedback
to an earlier version of this manuscript.
We also thank Grant Lythe
and Mart\'in L\'opez-Garc\'ia for
their input in some of the early
ideas behind this work and for supervising
the doctoral project of one of us (LS).
This manuscript has been internally reviewed at Los Alamos National Laboratory, and assigned
the reference number
LA-UR-22-25806 (CMP).



\bibliographystyle{unsrt}
\bibliography{refs}

\end{document}